\documentclass[a4paper,11pt]{article}

\pdfoutput=1

\usepackage{amsmath,amssymb,mathtools}
\usepackage{color}
\usepackage{graphicx}
\usepackage{subfigure}
\usepackage{cite}
\usepackage[colorlinks=true,linkcolor=red,citecolor=blue,urlcolor=blue,bookmarks]{hyperref}
\usepackage{multirow,makecell}
\usepackage{textcomp}
\usepackage{wasysym}
\usepackage{ulem}

\usepackage{verbatim}

\usepackage[utf8]{inputenc}
\usepackage[T1]{fontenc}

\usepackage[text={17cm,25cm},centering]{geometry} 

\numberwithin{equation}{section}

\def \be {\begin{equation}}
\def \ee {\end{equation}}
\def \ba {\begin{array}}
\def \ea {\end{array}}
\def \bea {\begin{eqnarray}}
\def \eea {\end{eqnarray}}
\def \nn {\nonumber}

\def \a {\alpha}

\def \d {\delta}

\def \ve {\varepsilon}

\def \s {\sigma}

\def \r {\rho}

\def \th {\theta}

\def \io {\iota}

\def \cH {\mathcal H}
\def \cI {\mathcal I}

\def \cM {\mathcal M}
\def \cN {\mathcal N}

\def \cU {\mathcal U}

\def \cX {\mathcal X}

\def \f {\frac}

\def \sr {\sqrt}
\def \td {\tilde}

\def \inf {\infty}

\def \lag {\langle}
\def \rag {\rangle}

\def \dd {\mathrm{d}}
\def \ep {\mathrm{e}}
\def \ii {\mathrm{i}}

\def \ves {\varnothing}

\def \tr {\textrm{tr}}

\def \and {{~\textrm{and}~}}

\def \cl {{\textrm{cl}}}

\def \SSH {{\textrm{SSH}}}

\def \univ {{\textrm{univ}}}

\def \bos {{\textrm{bos}}}
\def \fer {{\textrm{fer}}}
\def \XXX {{\textrm{XXX}}}
\def \soft {{\textrm{soft}}}
\def \hard {{\textrm{hard}}}

\def \I {{\textrm{I}}}
\def \II {{\textrm{II}}}
\def \IIIa {{\textrm{IIIa}}}
\def \IIIb {{\textrm{IIIb}}}

\begin{document}

\title{
\textbf{Shannon entropy in quasiparticle states of quantum chains}
}
\author{
Wentao Ye$^{1}$
and
Jiaju Zhang$^{1,2}$\footnote{Corresponding author: jiajuzhang@tju.edu.cn}
}
\date{}
\maketitle
\vspace{-10mm}
\begin{center}
{\it
$^{1}$Department of Physics, School of Science, Tianjin University,\\
      135 Yaguan Road, Tianjin 300350, China\\\vspace{1mm}
$^{2}$Center for Joint Quantum Studies, School of Science, Tianjin University,\\
      135 Yaguan Road, Tianjin 300350, China
}
\vspace{10mm}
\end{center}

\begin{abstract}

  We investigate the Shannon entropy of the total system and its subsystems, as well as the subsystem Shannon mutual information, in quasiparticle excited states of free bosonic and fermionic chains and the ferromagnetic phase of the spin-1/2 XXX chain. For single-particle and double-particle states, we derive various analytical formulas for free bosonic and fermionic chains in the scaling limit. These formulas are also applicable to certain magnon excited states in the XXX chain in the scaling limit. We also calculate numerically the Shannon entropy and mutual information for triple-particle and quadruple-particle states in bosonic, fermionic, and XXX chains. We discover that Shannon entropy, unlike entanglement entropy, typically does not separate for quasiparticles with large momentum differences. Moreover, in the limit of large momentum difference, we obtain universal quantum bosonic and fermionic results that are generally distinct and cannot be explained by a semiclassical picture.

\end{abstract}

\baselineskip 18pt
\thispagestyle{empty}
\newpage


\tableofcontents

\section{Introduction}

Quasiparticles are interesting collective excitations in integrable many-body systems that often provide simple intuitive explanations for complex phenomena \cite{Venema:2016rla}. One such example is the entanglement entropy in quasiparticle states of integrable quantum spin chains, which displays intriguing universal features in the scaling limit \cite{Pizorn:2012aut,Berkovits:2013mii,Molter:2014qsb,Castro-Alvaredo:2018dja,Castro-Alvaredo:2018bij,Castro-Alvaredo:2019irt,%
Castro-Alvaredo:2019lmj,Zhang:2020ouz,Zhang:2020vtc,Zhang:2020dtd,Zhang:2020txb,Zhang:2021bmy,Mussardo:2021gws,Capizzi:2022jpx,Capizzi:2022nel,%
Capizzi:2023ksc}. It was found that under certain limits, the entanglement entropy in quasiparticle states displays universal behaviors that can be explained by a semiclassical quasiparticle picture \cite{Castro-Alvaredo:2018dja,Castro-Alvaredo:2018bij}. The aim of the paper is to investigate whether there exists a similar semiclassical picture for other quantities, such as the total system Shannon entropy, subsystem Shannon entropy, and subsystem Shannon mutual information, in excited states of quasiparticles, akin to those observed for entanglement entropy.

In a state that is pure, the quantum correlation between a subsystem and its complement could be described by various entanglement measures, including the entanglement entropy, which refers to the von Neumann entropy of the reduced density matrix.
Depending on the particular quantum state of a given many-body system, the entanglement entropy exhibits various behaviors in the scaling limit \cite{Amico:2007ag,Eisert:2008ur,Calabrese:2009bph,Laflorencie:2015eck,Witten:2018lha}.
In this paper, we restrict our analysis to the bipartite case, where the entire system is divided intoa subsystem $A$ and its complement $\bar{A}$.
A quasiparticle state could be represented by the momenta $K$ of the excited quasiparticles, denoted as $|K\rangle$.
The reduced density matrix of subsystem $A$ can be denoted as $\rho_A=\tr_{\bar A}|K\rag\lag K|$, and the entanglement entropy can be calculated as $S_{A,K}=-\tr_A(\rho_A\log\rho_A)$.
It has been discovered that under the large energy condition
\be \label{condition1}
\max_{k \in K} \f{1}{\ve_k} \ll \min(\ell,L-\ell),
\ee
where $\ve_k$ is energy of the elementary excitation of one quasiparticle with momentum $k$, and the large momentum difference condition
\be \label{condition2}
|k-k'| \gg 1, ~ \forall k \in K, \forall k'\in K',
\ee
the entanglement entropy difference is given by \cite{Zhang:2021bmy}
\be \label{EED}
S_{A,K \cup K'} - S_{A,K'} = - \tr_A( \td \r_{A,K} \log \td \r_{A,K} ),
\ee
where $\td \r_{A,K}$ is an effective low-rank reduced density matrix.
In other words, under the conditions (\ref{condition1}) and (\ref{condition2}), the spin chain with modes $K'$ acts like a background, and contributions from the modes $K$ to the entanglement entropy decouple from the background.
Analytical formulas of the entanglement entropy difference in free bosonic and fermionic chains were obtained in \cite{Zhang:2020vtc,Zhang:2020dtd,Zhang:2021bmy}, and these formulas and their proper combinations also apply to models with interactions such as spin-1/2 XXX chain and XXZ chain under certain conditions \cite{Zhang:2021bmy}.
Furthermore, under the extra large momentum difference condition \cite{Zhang:2020vtc,Zhang:2020dtd}
\be
|k_1-k_2| \gg 1, ~ \forall k_1,k_2 \in K, k_1\neq k_2,
\ee
the effective reduced density matrix $\td \r_{A,K}$ has a simple semiclassical quasiparticle picture and the entanglement entropy difference (\ref{EED}) becomes the Shannon entropy of the probability distribution of the semiclassical quasiparticles \cite{Castro-Alvaredo:2018dja,Castro-Alvaredo:2018bij}.
It was found recently that similar universal properties and semiclassical quasiparticle picture also apply to the subsystem distance in quasiparticle excited states of various quantum spin chains \cite{Zhang:2020ouz,Zhang:2020txb,Zhang:2022tgu}.

In quantum mechanics, one can express the state of a quantum system in a pure state $|\psi\rangle$ in any orthonormal basis $\{|i\rangle\}$ with $\langle i|j\rangle=\delta_{ij}$ as
\be
|\psi\rangle = \sum_i c_i |i\rangle,
\ee
where the normalization of the state $\langle\psi|\psi\rangle=1$ implies that $\{p_i=|c_i|^2\}$ forms a well-defined probability distribution, i.e., $p_i\geq0$ and $\sum_i p_i=1$. A quantum system in a mixed state is characterized by a density matrix $\rho$, which is positive semi-definite and satisfies $\tr\rho=1$. One can write the density matrix in the orthonormal basis $\{|i\rangle\}$ as
\be
\rho = \sum_{i,j} \rho_{ij} |i\rangle\langle j|,
\ee
and there is a well-defined probability distribution $\{p_i=\rho_{ii}\}$. The Shannon entropy \cite{Nielsen:2010oan,Watrous:2018rgz}
\be
H=-\sum_i p_i \log p_i,
\ee
can be calculated for any arbitrary probability distribution $\{p_i\}$, which measures the uncertainty or randomness of the probability distribution. In the case of a quantum system in a pure or mixed state, the Shannon entropy depends on the chosen basis. In quantum spin chains, the direct product of the local basis at each individual site forms a natural basis. In the local basis, the Shannon entropy characterizes a type of interesting correlation between different parts of the quantum system.
The Shannon entropy in the local basis of various quantum spin chains has been intensively studied for both the total system %
\cite{Wolf:2007tdq,Stephan:2009gsi,Stephan:2010sba,Luitz:2013rga,Luitz:2014tah,Luitz:2014oaq,Dong:2022fqg} and a subsystem %
\cite{Lau:2012yvf,Stephan:2013efn,Alcaraz:2013ara,Stephan:2014nda,Alcaraz:2014tfa,Alcaraz:2015yea,Getelina:2015mko,Najafi:2015rma,%
Alcaraz:2016xaw,Tarighi:2022utu}.
The probabilities of finding different states in this basis are called formation probabilities \cite{Najafi:2015rma}. These probabilities, especially the emptiness formation probability, have been the subject of extensive research %
\cite{Korepin:1994ui,Essler:1994se,Essler:1995vp,Shiroishi:2001bji,Kitanine:2002vkb,Korepin:2002wyq,Franchini:2005uv,Stephan:2010sba,%
Stephan:2013efn,Najafi:2015rma,Rajabpour:2015hnr,Rajabpour:2016iyh,Ares:2019rad,Najafi:2019ypm,Ares:2020uwy,Ares:2020ghj}.
The Shannon entropy, which is dependent on the choice of basis, is more experimentally accessible than the basis-independent entanglement entropy. Measuring entanglement entropy in a general quantum state often necessitates the complex process of quantum state tomography \cite{James:2001owq}, which can be challenging. In contrast, determining the Shannon entropy involves performing measurements in a chosen basis on multiple copies of the quantum state to obtain the probability distribution of outcomes, which can then be used to calculate the Shannon entropy.

\begin{figure}[t]
  \centering
  \includegraphics[height=0.2\textwidth]{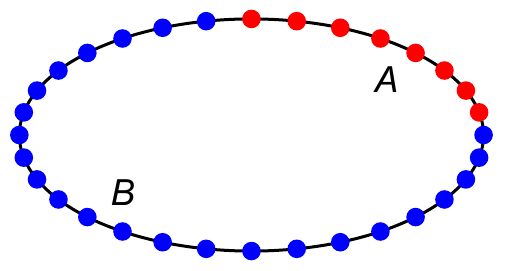}\\
  \caption{The spin chain consists of $L$ sites arranged in a circle. We consider a connected subsystem $A$ with $\ell$ adjacent sites, defined as $A=[1,\ell]$, and its complement $B=[\ell+1,L]$.}
  \label{FigureCC}
\end{figure}

In this paper, we examine the Shannon entropy of both the entire system and a connected subsystem in states of quasiparticle excitations of free bosonic and fermionic chains and the spin-1/2 XXX chain.
We focus on translation invariant states, ensuring that the subsystem Shannon entropy remains independent of its position within the entire system and solely dependent on the subsystem size.
We not only consider the single-particle and double-particle states, which facilitate analytical calculations, but also the states with more particles, for which we calculate the Shannon entropy using numerical methods.
The results are then compared across diverse quantum spin chains and also with classical results. As depicted in Figure~\ref{FigureCC}, we consider a connected subsystem $A=[1,\ell]$ with $\ell$ neighboring sites on a circular chain consisting of $L$ sites. Only translation-invariant states in quantum spin chains and translation-invariant configurations in classical chains are taken into account. We compute the Shannon entropy of the entire system $H(L)$ and the subsystem Shannon entropy $H(\ell)$. Additionally, we evaluate the subsystem Shannon mutual information,
\be
M(\ell) = H(\ell) + H(L-\ell) - H(L),
\ee
which is a measure of the correlation between the subsystem $A$ and its complement $B$. Our focus is on understanding the behaviours in the total system Shannon entropy $H(L)$, the subsystem Shannon entropy $H(\ell)$, and the subsystem mutual information $I(\ell)$ as we approach the scaling limit $L\to+\infty$, $\ell\to+\infty$ while maintaining a constant ratio $x=\ell/L$.
As there is the semiclassical picture for the entanglement entropy only in the large momentum difference limit \cite{Zhang:2020vtc,Zhang:2020dtd}, we also take such a limit for the local basis Shannon entropy and see if there is a similar semiclassical picture.

In this paper, we present analytical formulas for the total and subsystem Shannon entropy and mutual information in single- and double-quasiparticle excited states of free bosonic and fermionic chains, as well as in single- and double-magnon excited states of the ferromagnetic phase of the spin-1/2 XXX chain.
We also calculated numerically the Shannon entropy and mutual information in triple-particle and quadruple-particle states in the bosonic, fermionic and XXX chain.
In the scaling limit where $L$ and $\ell$ approach infinity with a fixed ratio between subsystem size and total system size $x={\ell}/{L}$, both the total system Shannon entropy and subsystem Shannon entropy follow a universal logarithmic law, while the subsystem mutual information is a finite function of the ratio $x$.
The results obtained for free bosonic and fermionic chains can also be applied to the XXX chain under certain circumstances.
We compare our results with those of classical particles.
In a single-particle state, the results are trivial and universal.
In a general multi-particle state with large momentum differences, we find distinct universal bosonic and fermionic formulas that cannot be reproduced semiclassically.
In the scaling limit, the contributions from different classical particles decouple, whereas this is not the case for quantum quasiparticles, even in the large momentum difference limit.

The rest of this paper is organized in the following way. In sections~\ref{sectionFBC} and \ref{sectionFFC}, we present calculations of the total system Shannon entropy and subsystem Shannon entropy and mutual information in quasiparticle excited states in free bosonic and fermionic chains, respectively.
In section~\ref{sectionXXX}, we consider magnon excited states in the spin-1/2 XXX chain.
In section~\ref{sectionthreenfour}, we evaluate numerically the Shannon entropy and mutual information in the triple-particle and quadruple-particle states in the bosonic, fermionic and XXX chains.
The paper concludes with discussions in section~\ref{sectionConclusion}.
We collect the calculation details for free bosoic chain, free fermionic chain, and XXX chain in respectively appendices~\ref{appendixBoson}, \ref{appendixFermion} and \ref{appendixXXX}.
In appendix~\ref{appendixSSH}, we discuss briefly the entanglement entropy and Shannon entropy in the Su-Schrieffer-Heeger (SSH) model.
In appendix~\ref{appendixCP}, we consider the configurations of classical particles.
In appendix~\ref{appendixPNPD}, we calculate the Shannon entropy for the probability distribution of subsystem particle numbers in free bosonic and fermionic chains.
Finally, in appendix~\ref{appendixXB}, we study the Shannon entropy in the local basis of $\s_j^x$ eigenstates of the XXX chain.

\section{Free bosonic chain} \label{sectionFBC}

In this section, we evaluate the total system and subsystem Shannon entropies and mutual information for the free bosonic chain in the single-particle state $|k\rag$ and the double-particle states $|k^2\rag$ and $|k_1k_2\rag$.

The quasiparticle states analyzed in this paper, within the context of free models, can be derived by considering the limit of infinite energy gap from the eigenstates of nearest-neighbor interacting models. In essence, free models can be regarded as particular instances of interacting models. The conclusions presented in this section, particularly the absence of semiclassical representations for the local basis Shannon entropy and mutual information, and the inability to isolate contributions from quasiparticles with large momentum difference, are valid for both the free bosonic chain discussed and the subsequent free fermionic chain. These findings also extend to their corresponding nearest-neighbor interacting models.

The construction of the quasiparticle states and the corresponding local configuration probabilities are shown in appendix~\ref{appendixBoson}.

\subsection{Single-particle state $|k\rag$}\label{sectionbosk}

In the state $|k\rag$ with a single quasiparticle with momentum $k$, we get the Shannon entropy of the total system
\be
H^\bos_{k}(L) = \log L,
\ee
and the subsystem Shannon entropy
\be
H^\bos_{k}(\ell) = x \log L - (1-x) \log (1-x),
\ee
and the subsystem Shannon mutual information
\be
M_k^\bos(\ell) = - x \log x - (1-x) \log (1-x).
\ee

The results for the single-particle state $|k\rag$ are the same with the results for the configuration of one classical particle (\ref{Hcl1L}), (\ref{Hcl1ell}) and (\ref{Icl1ell}).

\subsection{Double-particle state $|k^2\rag$} \label{subsectiondpsk2}

We consider the state $|k^2\rag$ with the mode of momentum $k$ being excited twice.
In the scaling limit, we get the Shannon entropy
\be
H^\bos_{k^2}(L) = 2\log L-\log 2.
\ee
the subsystem Shannon entropy
\be
H^\bos_{k^2}(\ell) = 2x \log L - x(2-x) \log2 -2(1-x)\log(1-x),
\ee
and mutual information
\be
M_{k^2}^\bos(\ell)=-x^2\log x^2- 2x(1-x)\log [2x(1-x)]-(1-x)^2\log (1-x)^2.
\ee

The results for the double-particle state $|k^2\rag$ are the same as the results for the configuration of two identical classical particles (\ref{Hcl1hat2L}), (\ref{Hcl1hat2ell}), and (\ref{Icl1hat2ell}).
It is easy to check that the results in the general $r$-particle state $|k^r\rag=\f{1}{\sr{r!}}(b_k^\dag)^r|G\rag$ are the same as the results for the configuration of $r$ identical classical particles (\ref{Hcl1hatrL}), (\ref{Hcl1hatrell}), and (\ref{Icl1hatrell}). We skip the details of the calculation in this paper.

\subsection{Double-particle state $|k_1k_2\rag$}

In this subsection we calculate the Shannon entropy for the double-particle state $|k_1k_2\rag=b^\dag_{k_1}b^\dag_{k_2}|G\rag$ in free bosonic chain.

\subsubsection{Total system Shannon entropy}

For general $L$ and the momentum difference $k_{12}=k_1-k_2$, we get the Shannon entropy of the total system
\be \label{Hbosk1k2Lgeneral}
H^\bos_{k_1k_2}(L) = 2\log L - 2\log2 + \f{2\log2}{L}
                   - \f{4}{L} \sum_{j=1}^{L/2-1} \cos^2\f{\pi j k_{12}}{L} \log \cos^2\f{\pi j k_{12}}{L}.
\ee
For $|k_{12}| \ll L$ in the limit $L \to +\inf$, the Shannon entropy becomes
\be \label{Hbosk1k2Lk12llL}
H^\bos_{k_1k_2}(L) = H^\univ_{k_1k_2}(L) = 2\log L - 1, 
\ee
which does not depend on the actual values of the momenta $k_1,k_2$.
As we will see in the subsequent section, the formula (\ref{Hbosk1k2Lk12llL}) also applies to the free fermionic chain under the condition $|k_{12}| \ll L$.
So we also call it a universal result.

When $|k_{12}|$ is proportional to $L$ in the scaling limit $L \to +\inf$, the Shannon entropy may take exceptional values for exceptional values of $|k_{12}|$.
For $|k_{12}|=\f{mL}{n} \leq \f{L}2$ with the integer $n=2,3,4,\cdots$ being a divisor of $L$ and the integer $m$ being coprime with $n$, we get the total system Shannon entropy
\be \label{Hbosk1k2Lk12eqmLn}
\cH^\bos_{k_1k_2}(L) = 2\log L - 2\log2
                   - \f{2}{n} \sum_{a=1}^{n-1} \cos^2\f{\pi a}{n} \log \cos^2\f{\pi a}{n},
\ee
which is independent of $m$.
Explicitly, for $n=\{2,3,4,5,6\}$, we get respectively
\bea \label{EBR}
&& \cH^\bos_{k_1k_2}(L) - 2\log L =
\Big\{
         -2 \log 2,
         -\f43 \log 2,
         -\f32 \log 2, \nn \\
&& \phantom{\cH^\bos_{k_1k_2}(L) - 2\log L =}
         -\f15 \log 2 - \f{1}{10}[ (3-\sqrt{5}) \log (3-\sqrt{5}) + (3+\sqrt{5}) \log (3+\sqrt{5})  ], \nn \\
&& \phantom{\cH^\bos_{k_1k_2}(L) - 2\log L =}
        -\f23 \log 2 - \f12 \log 3
\Big\}.
\eea
For a prime integer $L$, there is no $|k_{12}|=\f{mL}{n} \leq \f{L}2$ with coprime integers $m,n$, and so the formula  (\ref{Hbosk1k2Lk12llL}) is valid for all $k_{12}$ for a large prime integer $L$.
Interestingly, in further $n\to+\inf$ limit, the exceptional formula (\ref{Hbosk1k2Lk12eqmLn}) approaches the universal formula (\ref{Hbosk1k2Lk12llL}).

\begin{figure}[t]
  \centering
  \includegraphics[height=0.28\textwidth]{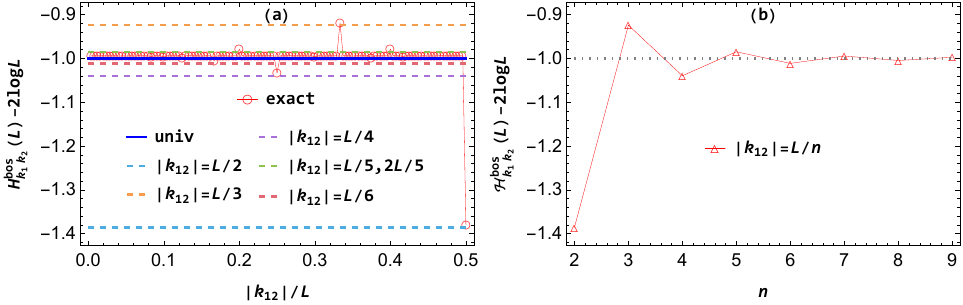}\\
  \caption{The Shannon entropy of the total system in the double-particle state $|k_1k_2\rag$ of the free bosonic chain.
  Left: The red empty circles connected with thin lines (exact) are the exact bosonic result (\ref{Hbosk1k2Lgeneral}) with $L=240$.
  The dark blue solid line (univ) is the result (\ref{Hbosk1k2Lk12llL}), i.e.\ the universal formula (\ref{Hbosk1k2Lk12llL}), which is valid for $|k_{12}| \ll L$.
  The dashed lines are the exceptional result (\ref{Hbosk1k2Lk12eqmLn}) which is valid for exceptional values of the momentum difference $|k_{12}|=\f{mL}{n} \leq \f{L}{2}$ with coprime integers $m,n$.
  Right: The large $n$ limit of exceptional bosonic result (\ref{Hbosk1k2Lk12eqmLn}) leads to the universal result (\ref{Hbosk1k2Lk12llL}).}
  \label{FigureHbosk1k2L}
\end{figure}

We have obtained three formulas for the total system Shannon entropy, i.e.\ the exact formula (\ref{Hbosk1k2Lgeneral}) written in terms of a summation which is valid for general $L$ and $k_{12}$, the formula (\ref{Hbosk1k2Lk12llL}), i.e.\ the universal formula (\ref{Hbosk1k2Lk12llL}), which is valid for $|k_{12}| \ll L$ in the scaling limit, and the exceptional formula (\ref{Hbosk1k2Lk12eqmLn}) which is valid for exceptional values of the momentum difference $|k_{12}|=\f{mL}{n} \leq \f{L}2$ with coprime integers $m,n$ in the scaling limit.
We show these results in Figure~\ref{FigureHbosk1k2L}.
In the left panel of Figure~\ref{FigureHbosk1k2L}, we see that for most values of momentum difference $|k_{12}|$, not only for $|k_{12}|\ll L$, the Shannon entropy of the total system is $(\ref{Hbosk1k2Lk12llL})$, i.e.\ the universal formula (\ref{Hbosk1k2Lk12llL}).
Only for a few exceptional values of $|k_{12}|=\f{mL}{n} \leq \f{L}2$ with coprime integers $m,n$, the Shannon entropy takes the exceptional form (\ref{Hbosk1k2Lk12eqmLn}).
In the right panel of Figure~\ref{FigureHbosk1k2L}, it is shown that the $n\to+\inf$ limit of exceptional result (\ref{Hbosk1k2Lk12eqmLn}) leads to the universal result (\ref{Hbosk1k2Lk12llL}).

\subsubsection{Subsystem Shannon entropy}

For general $L$, $\ell$, and $k_{12}$, we get the exact subsystem Shannon entropy
\bea \label{Hbosk1k2ellgeneral}
&& \hspace{-8mm}
   H^\bos_{k_1k_2}(\ell) = - \Big[ (1 - x)^2+\f{\sin^2(\pi k_{12} x)}{L^2\sin^2\f{\pi k_{12}}{L}} \Big]
                        \log \Big[ (1 - x)^2+\f{\sin^2(\pi k_{12} x)}{L^2\sin^2\f{\pi k_{12}}{L}} \Big] \nn\\
&& \hspace{-2mm} - \sum_{j=1}^\ell \Big[ \f{2(1-x)}{L} - \f{2\sin(\pi k_{12} x)\cos\f{2\pi k_{12} (j-\f{\ell+1}{2})}{L}}{L^2\sin\f{\pi k_{12}}{L}} \Big] \log  \Big[ \f{2(1-x)}{L} - \f{2\sin(\pi k_{12} x)\cos\f{2\pi k_{12} (j-\f{\ell+1}{2})}{L}}{L^2\sin\f{\pi k_{12}}{L}} \Big] \nn\\
&& \hspace{-2mm} + \f{2x}{L} ( 2\log L - \log 2)
   - \sum_{j=1}^{\ell-1} ( \ell-j ) \f{4}{L^2}\cos^2\f{\pi j k_{12}}{L} \log\Big( \f{4}{L^2}\cos^2\f{\pi j k_{12}}{L} \Big).
\eea
For $|k_{12}| \ll L$ in the scaling limit $L\to+\inf$, $\ell\to+\inf$ with fixed $x={\ell}/{L}$, we get the Shannon entropy
\bea \label{Hbosk1k2ellk12llL}
&& \hspace{-8mm}
   H^\bos_{k_1k_2}(\ell) = 2x\log L - 2x \log 2 - \Big[ (1 - x)^2+\f{\sin^2(\pi k_{12} x)}{\pi^2 k_{12}^2} \Big]
                                             \log \Big[ (1 - x)^2+\f{\sin^2(\pi k_{12} x)}{\pi^2 k_{12}^2} \Big]\nn\\
&& \hspace{-8mm}\phantom{H^\bos_{k_1k_2}(\ell) =}
   - 4 \int_0^{x/2} \!\! \dd y \Big[ (1-x) - \f{\sin(\pi k_{12} x)\cos(2\pi k_{12} y)}{\pi k_{12}} \Big]
                          \log \Big[ (1-x) - \f{\sin(\pi k_{12} x)\cos(2\pi k_{12} y)}{\pi k_{12}} \Big] \nn\\
&& \hspace{-8mm}\phantom{H^\bos_{k_1k_2}(\ell) =}
   -4 \int_0^{k_{12} x} \!\! \f{\dd z}{k_{12}} \Big( x-\f{z}{k_{12}} \Big) \cos^2(\pi z) \log \cos^2(\pi z).
\eea
For $1 \ll |k_{12}| \ll L$ in the scaling limit, we get the universal result
\be \label{Hunivk1k2ell}
H^\univ_{k_1k_2}(\ell) = 2x \log L - 2 (1-x) \log(1-x) -x^2 - 2x(1-x)\log 2,
\ee
which is not the same as either the result for two identical classical particles $H^\cl_{1^2}(\ell)$ (\ref{Hcl1hat2ell}) or the result for two distinguishable classical particles $H^\cl_{12}(\ell)$ (\ref{Hcl12ell}).
As we will see in the subsequent section, the universal result also applies to the double-particle state $|k_1k_2\rag$ in the free fermionic theory.

For exceptional values of the momentum difference $|k_{12}|=\f{mL}{n} \leq \f{L}{2}$ with coprime integers $m,n$ in the scaling limit $L\to+\inf$, $\ell\to+\inf$ with fixed $x={\ell}/{L}$, we get the exceptional value of the subsystem Shannon entropy
\be \label{Hbosk1k2ellk12eqmLn}
\cH^\bos_{k_1k_2}(\ell) = 2 x \log L - 2x \log 2 - 2(1-x)\log(1-x)
                      - \f{2x^2}{n} \sum_{a=1}^{n-1} \cos^2\f{\pi a}{n} \log \cos^2\f{\pi a}{n}.
\ee
Note that the $x\to1$ limit of (\ref{Hbosk1k2ellk12eqmLn}) is just (\ref{Hbosk1k2Lk12eqmLn}) as it should be.
Also, the $n\to+\inf$ limit of the exceptional result (\ref{Hbosk1k2ellk12eqmLn}) leads to the universal result (\ref{Hunivk1k2ell}).

\begin{figure}[t]
  \centering
  \includegraphics[height=0.28\textwidth]{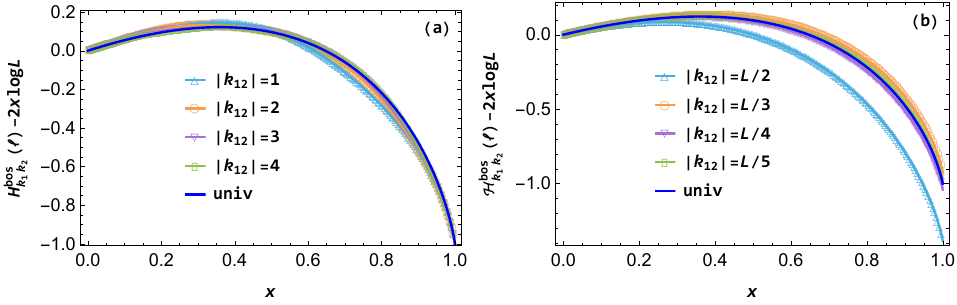}\\
  \caption{The subsystem Shannon entropy in the double-particle state $|k_1k_2\rag$ of the free bosonic chain.
  The empty symbols are the exact result (\ref{Hbosk1k2ellgeneral}) with $L=240$.
  The dark blue solid lines are the universal result (\ref{Hunivk1k2ell}) which is valid for $1 \ll |k_{12}| \ll L$.
  In the left panel the other solid lines are the result (\ref{Hbosk1k2ellk12llL}) which is valid for $|k_{12}| \ll L$, and in the right panel the other solid lines are the exceptional result (\ref{Hbosk1k2ellk12eqmLn}) which is valid for exceptional values $|k_{12}|=\f{mL}{n} \leq \f{L}{2}$ with coprime integers $m,n$.}
  \label{FigureHbosk1k2ell}
\end{figure}

In summary, we have obtained four formulas for the subsystem Shannon entropy, i.e.\
the exact formula (\ref{Hbosk1k2ellgeneral}) which is valid for general $L$, $\ell$, and $k_{12}$,
the formula (\ref{Hbosk1k2ellk12llL}) which is valid for $|k_{12}| \ll L$ in the scaling limit $L\to+\inf$, $\ell\to+\inf$ with fixed $x={\ell}/{L}$,
the universal formula (\ref{Hunivk1k2ell}) which is valid for $1 \ll |k_{12}| \ll L$ in the scaling limit,
the exceptional formula (\ref{Hbosk1k2ellk12eqmLn}) which is valid for the exceptional values $|k_{12}|=\f{mL}{n} \leq \f{L}{2}$ with coprime integers $m,n$ in the scaling limit.
We show these results in Figure~\ref{FigureHbosk1k2ell}, where in the left panel we see the large $|k_{12}|$ limit of the bosonic result (\ref{Hbosk1k2ellk12llL}), which gives the universal result (\ref{Hunivk1k2ell}), and in the right panel we see the large $n$ limit of the exceptional result (\ref{Hbosk1k2ellk12eqmLn}), which also gives the universal result (\ref{Hunivk1k2ell}).

\subsubsection{Subsystem Shannon mutual information}

\begin{figure}[t]
  \centering
  \includegraphics[height=0.28\textwidth]{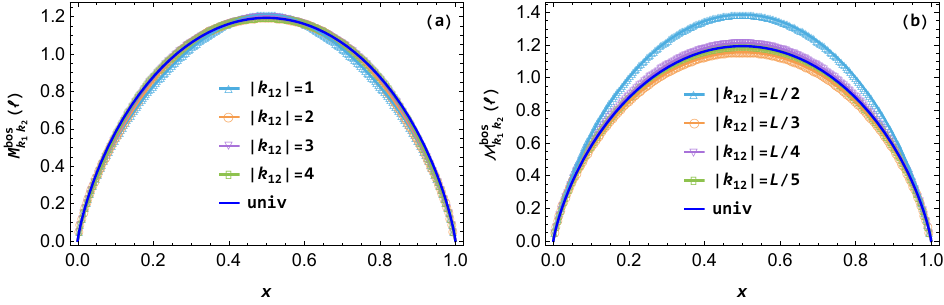}\\
  \caption{The subsystem Shannon mutual information in the double-particle state $|k_1k_2\rag$ of the free bosonic chain.
  The empty symbols are the exact result with $L=240$.
  The dark blue solid lines are the universal result (\ref{Iunivk1k2ell}) which is valid for $1 \ll |k_{12}| \ll L$.
  In the left panel the other solid lines are the result which is valid for $|k_{12}| \ll L$, and in the right panel the other solid lines are the exceptional result (\ref{Ibosk1k2ellk12eqmLn}) which is valid for the exceptional values $|k_{12}|=\f{mL}{n} \leq \f{L}{2}$ with coprime integers $m,n$.}
  \label{FigureMbosk1k2ell}
\end{figure}

Using the results of the total system and subsystem Shannon entropies in different parameter regimes, we evaluate the subsystem Shannon mutual information.
From (\ref{Hbosk1k2Lgeneral}) and (\ref{Hbosk1k2ellgeneral}), we obtain the exact mutual information $M^\bos_{k_1k_2}(\ell)$ that is valid for general $L$, $\ell$, $k_{12}$.
From (\ref{Hbosk1k2Lk12llL}) and (\ref{Hbosk1k2ellk12llL}), we obtain the mutual information
\be \label{Ibosk1k2ellk12llL}
M^\bos_{k_1k_2}(\ell) = H^\bos_{k_1k_2}(\ell) + H^\bos_{k_1k_2}(L-\ell) - H^\bos_{k_1k_2}(L),
\ee
which is valid for $|k_{12}| \ll L$ in the scaling limit.
From (\ref{Hbosk1k2Lk12llL}) and (\ref{Hunivk1k2ell}), we further obtain the universal mutual information
\be \label{Iunivk1k2ell}
M^\univ_{k_1k_2}(\ell) = -2x\log{x}-2(1-x)\log{(1-x)} -2x(1-x)(2\log{2}-1),
\ee
which is valid for $1\ll |k_{12}| \ll L$ in the scaling limit.
From (\ref{Hbosk1k2Lk12eqmLn}) and (\ref{Hbosk1k2ellk12eqmLn}), we obtain the exceptional mutual information
\be \label{Ibosk1k2ellk12eqmLn}
\cM^\bos_{k_1k_2}(\ell) = -2x\log{x}-2(1-x)\log{(1-x)} + \f{4x(1-x)}{n} \sum_{a=1}^{n-1} \cos^2\f{\pi a}{n} \log \cos^2\f{\pi a}{n},
\ee
which is valid for the exceptional values $|k_{12}|=\f{mL}{n} \leq \f{L}{2}$ with coprime integers $m,n$.
In the scaling limit, the Shannon mutual information is a finite function of the ratio $x={\ell}/{L}$.
Surprisingly, both the results (\ref{Iunivk1k2ell}) and (\ref{Ibosk1k2ellk12eqmLn}) are generally different from the result for two distinguishable classical particles $M^\cl_{12}(\ell)$ (\ref{Icl12ell}).
Only for the special case with $|k_{12}|=\f{L}{2}$, there is $\cM^\bos_{k_1k_2}(\ell)=M^\cl_{12}(\ell)$.
The results (\ref{Iunivk1k2ell}) and (\ref{Ibosk1k2ellk12eqmLn}) are also different from the result for two identical classical particles $M^\cl_{1^2}(\ell)$ (\ref{Icl1hat2ell}).
We show the subsystem Shannon mutual information in Figure~\ref{FigureMbosk1k2ell}.

\section{Free fermionic chain} \label{sectionFFC}

This section deals with the single-particle state $|k\rag$ and double-particle state $|k_1k_2\rag$ in the free fermionic chain.
The calculation details are collected in appendix~\ref{appendixFermion}.
The case of the free fermionic chain resembles that of the free bosonic chain, so we will keep it brief in this section.

\subsection{Single-particle state $|k\rag$} \label{sectionferk}

We can use the same calculations and results as for the single particle state $|k\rag$ in the free bosonic chain in subsection~\ref{sectionbosk}. We will skip them here.

\subsection{Double-particle state $|k_1k_2\rag$}

This subsection deals with the calculation of Shannon entropy in the double-particle state $|k_1k_2\rag=b^\dag_{k_1}b^\dag_{k_2}|G\rag$ in free fermionic chain.

\subsubsection{Total system Shannon entropy}

For general $L$ and $k_{12}$, we get the Shannon entropy of the total system
\be \label{Hferk1k2Lgeneral}
H^\fer_{k_1k_2}(L) = 2\log L - 2\log2
                   - \f{4}{L} \sum_{j=1}^{L/2-1} \sin^2\f{\pi j k_{12}}{L} \log \sin^2\f{\pi j k_{12}}{L}.
\ee
For $|k_{12}| \ll L$ in the limit $L \to +\inf$, the Shannon entropy becomes
\be \label{Hferk1k2Lk12llL}
H^\fer_{k_1k_2}(L) = 2\log L - 1,
\ee
which is the same as the bosonic result $H^\bos_{k_1k_2}(L)$ and the universal result $H^\univ_{k_1k_2}(L)$ in (\ref{Hbosk1k2Lk12llL}).
This is why we call the expression a universal result.

For the exceptional value $|k_{12}|=\f{mL}{n} \leq \f{L}2$ with coprime integers $m,n$, we get the exceptional Shannon entropy
\be \label{Hferk1k2Lk12eqmLn}
\cH^\fer_{k_1k_2}(L) = 2\log L - 2\log2
                   - \f{2}{n} \sum_{a=1}^{n-1} \sin^2\f{\pi a}{n} \log \sin^2\f{\pi a}{n}.
\ee
Explicitly, for $n=\{2,3,4,5,6\}$, we get respectively
\bea
&& \cH^\fer_{k_1k_2}(L) - 2\log L =
\Big\{
         -2 \log 2,
         - \log 3,
         -\f32 \log 2, \nn \\
&& \phantom{\cH^\fer_{k_1k_2}(L) - 2\log L =}
         \log 2 - \f{1}{10}[ (5-\sqrt{5}) \log (5-\sqrt{5}) + (5+\sqrt{5}) \log (5+\sqrt{5})  ], \nn \\
&& \phantom{\cH^\fer_{k_1k_2}(L) - 2\log L =}
        -\f23 \log 2 - \f12 \log 3
\Big\},
\eea
which are generally different from the exceptional bosonic results (\ref{EBR}). Again, in further large $n$ limit, the exceptional result approaches the universal result.

\begin{figure}[t]
  \centering
  \includegraphics[height=0.28\textwidth]{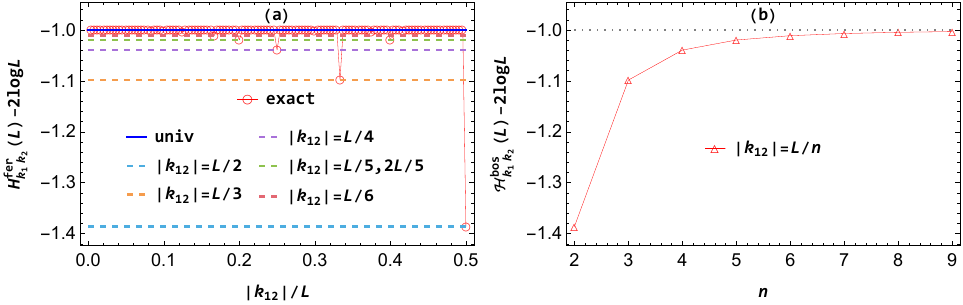}\\
  \caption{The Shannon entropy of the total system in the double-particle state $|k_1k_2\rag$ of the free fermionic chain.
  Left: The red empty circles connected with thin lines (exact) are the exact result (\ref{Hferk1k2Lgeneral}) with $L=240$.
  The dark blue solid line (univ) is the result (\ref{Hferk1k2Lk12llL}), i.e.\ the universal formula (\ref{Hbosk1k2Lk12llL}), which is valid for $|k_{12}| \ll L$.
  The dashed lines are the exceptional result (\ref{Hferk1k2Lk12eqmLn}) which is valid for exceptional values of the momentum difference $|k_{12}|=\f{mL}{n} \leq \f{L}{2}$ with coprime integers $m,n$.
  Right: The large $n$ limit of the exceptional result (\ref{Hferk1k2Lk12eqmLn}) leads to the universal result (\ref{Hbosk1k2Lk12llL}).}
  \label{FigureHferk1k2L}
\end{figure}

We show these various results in Figure~\ref{FigureHferk1k2L}.

\subsubsection{Subsystem Shannon entropy}

For general $L$, $\ell$, and $k_{12}$, we get the exact subsystem Shannon entropy
\bea \label{Hferk1k2ellgeneral}
&& \hspace{-8mm}
   H^\fer_{k_1k_2}(\ell) = - \Big[ (1 - x)^2 - \f{\sin^2(\pi k_{12} x)}{L^2\sin^2\f{\pi k_{12}}{L}} \Big]
                        \log \Big[ (1 - x)^2 - \f{\sin^2(\pi k_{12} x)}{L^2\sin^2\f{\pi k_{12}}{L}} \Big] \nn\\
&& - \sum_{j=1}^\ell \Big[ \f{2(1-x)}{L} + \f{2\sin(\pi k_{12} x)\cos\f{2\pi k_{12} (j-\f{\ell+1}{2})}{L}}{L^2\sin\f{\pi k_{12}}{L}} \Big] \log  \Big[ \f{2(1-x)}{L} + \f{2\sin(\pi k_{12} x)\cos\f{2\pi k_{12} (j-\f{\ell+1}{2})}{L}}{L^2\sin\f{\pi k_{12}}{L}} \Big] \nn\\
&& 
   - \sum_{j=1}^{\ell-1} ( \ell-j ) \f{4}{L^2}\sin^2\f{\pi j k_{12}}{L} \log\Big( \f{4}{L^2}\sin^2\f{\pi j k_{12}}{L} \Big).
\eea
For $|k_{12}| \ll L$ in the scaling limit $L\to+\inf$, $\ell\to+\inf$ with fixed $x={\ell}/{L}$, we get the Shannon entropy
\bea \label{Hferk1k2ellk12llL}
&& \hspace{-8mm}
   H^\fer_{k_1k_2}(\ell) = 2x\log L - 2x \log 2 - \Big[ (1 - x)^2 - \f{\sin^2(\pi k_{12} x)}{\pi^2 k_{12}^2} \Big]
                                             \log \Big[ (1 - x)^2 - \f{\sin^2(\pi k_{12} x)}{\pi^2 k_{12}^2} \Big]\nn\\
&& \hspace{-8mm}\phantom{H^\fer_{k_1k_2}(\ell) =}
   - 4 \int_0^{x/2} \!\! \dd y \Big[ (1-x) + \f{\sin(\pi k_{12} x)\cos(2\pi k_{12} y)}{\pi k_{12}} \Big]
                          \log \Big[ (1-x) + \f{\sin(\pi k_{12} x)\cos(2\pi k_{12} y)}{\pi k_{12}} \Big] \nn\\
&& \hspace{-8mm}\phantom{H^\fer_{k_1k_2}(\ell) =}
   -4 \int_0^{k_{12} x} \!\! \f{\dd z}{k_{12}} \Big( x-\f{z}{k_{12}} \Big) \sin^2(\pi z) \log \sin^2(\pi z).
\eea
For $1 \ll k_{12} \ll L$ in the scaling limit, we get the universal result $H^\univ_{k_1k_2}(\ell)$ (\ref{Hunivk1k2ell}).

For exceptional values of the momentum difference $|k_{12}|=\f{mL}{n} \leq \f{L}{2}$ with coprime integers $m,n$ in the scaling limit $L\to+\inf$, $\ell\to+\inf$ with fixed $x={\ell}/{L}$, we get the subsystem Shannon entropy
\be \label{Hferk1k2ellk12eqmLn}
\cH^\fer_{k_1k_2}(\ell) = 2 x \log L - 2x \log 2 - 2(1-x)\log(1-x)
                      - \f{2x^2}{n} \sum_{a=1}^{n-1} \sin^2\f{\pi a}{n} \log \sin^2\f{\pi a}{n}.
\ee

\begin{figure}[t]
  \centering
  \includegraphics[height=0.28\textwidth]{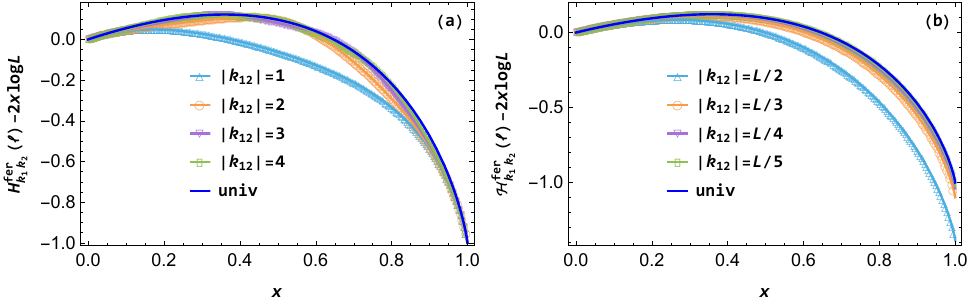}\\
  \caption{The subsystem Shannon entropy in the double-particle state $|k_1k_2\rag$ of the free fermionic chain.
  The empty symbols are the exact result (\ref{Hferk1k2ellgeneral}) with $L=240$.
  The dark blue solid lines are the universal result (\ref{Hunivk1k2ell}) which is valid for $1 \ll |k_{12}| \ll L$.
  In the left panel the other solid lines are the result (\ref{Hferk1k2ellk12llL}) which is valid for $|k_{12}| \ll L$, and in the right panel the other solid lines are the exceptional result (\ref{Hferk1k2ellk12eqmLn}) which is valid for the exceptional values $|k_{12}|=\f{mL}{n} \leq \f{L}{2}$ with coprime integers $m,n$.}
  \label{FigureHferk1k2ell}
\end{figure}

We show the various results in Figure~\ref{FigureHferk1k2ell}.

\subsubsection{Subsystem Shannon mutual information}

\begin{figure}[t]
  \centering
  \includegraphics[height=0.28\textwidth]{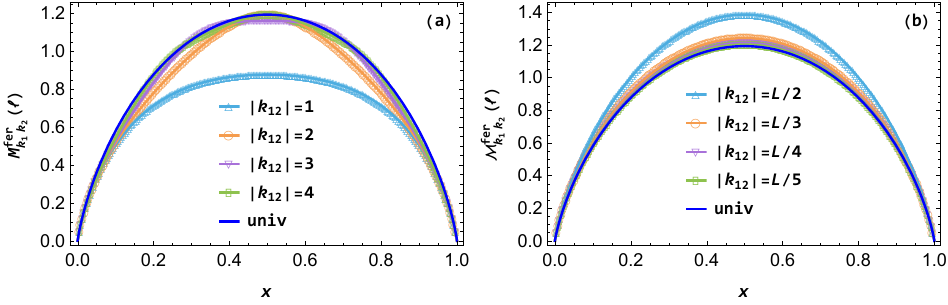}\\
  \caption{The subsystem Shannon mutual information in the double-particle state $|k_1k_2\rag$ of the free fermionic chain.
  The empty symbols are the exact result with $L=240$.
  The dark blue solid lines are the universal result (\ref{Iunivk1k2ell}) which is valid for $1 \ll |k_{12}| \ll L$.
  In the left panel the other solid lines are the result (\ref{Iferk1k2ellk12llL}) which is valid for $|k_{12}| \ll L$, and in the right panel the other solid lines are the exceptional result (\ref{Iferk1k2ellk12eqmLn}) which is valid for the exceptional values $|k_{12}|=\f{mL}{n} \leq \f{L}{2}$ with coprime integers $m,n$.}
  \label{FigureMferk1k2ell}
\end{figure}

With the results of the total system and subsystem Shannon entropies in different regimes of the parameters, we calculate the subsystem Shannon mutual information.
From (\ref{Hferk1k2Lgeneral}) and (\ref{Hferk1k2ellgeneral}), we obtain the exact fermionic mutual information $M^\fer_{k_1k_2}(\ell)$ that is valid for general $L$, $\ell$, $k_{12}$.
From (\ref{Hferk1k2Lk12llL}) and (\ref{Hferk1k2ellk12llL}), we obtain
\be \label{Iferk1k2ellk12llL}
M^\fer_{k_1k_2}(\ell) = H^\fer_{k_1k_2}(\ell) + H^\fer_{k_1k_2}(L-\ell) - H^\fer_{k_1k_2}(L),
\ee
which is valid for $|k_{12}| \ll L$ in the scaling limit.
From (\ref{Hferk1k2Lk12llL}) and (\ref{Hunivk1k2ell}), we obtain the universal mutual information $M^\univ_{k_1k_2}(\ell)$ which is the same as (\ref{Iunivk1k2ell}) and is valid for $1\ll |k_{12}| \ll L$ in the scaling limit.
From (\ref{Hferk1k2Lk12eqmLn}) and (\ref{Hferk1k2ellk12eqmLn}), we obtain the exceptional mutual information
\be \label{Iferk1k2ellk12eqmLn}
\cM^\fer_{k_1k_2}(\ell) = -2x\log{x}-2(1-x)\log{(1-x)} + \f{4x(1-x)}{n} \sum_{a=1}^{n-1} \sin^2\f{\pi a}{n} \log \sin^2\f{\pi a}{n},
\ee
which is valid for the exceptional values of the momentum difference $|k_{12}|=\f{mL}{n} \leq \f{L}{2}$ with coprime integers $m,n$.
We show the subsystem mutual information in Figure~\ref{FigureMferk1k2ell}.

\section{XXX chain}  \label{sectionXXX}

In this section, we compute the Shannon entropy of the total system and its subsystem and the subsystem mutual information in single- and double-magnon excited states of the spin-1/2 XXX chain. The single-magnon state results are identical to those for the single-particle states in free bosonic and fermionic chains. The double-magnon state is more complex, as it can be either a scattering state or a bound state. The total system Shannon entropy part in this section has overlaps with \cite{Dong:2022fqg}.

\subsection{Single-magnon state} \label{sectionXXXI}

In the single-magnon state $|I\rag$ total system Shannon entropy and the subsystem Shannon entropy and mutual information are the same as those of one-particle states in the free bosonic and fermionic chains in, respectively, subsections~\ref{sectionbosk} and \ref{sectionferk} and the configurations of one soft-core and hard-core classical particles in, respectively, subsections~\ref{sectionsoft1} and \ref{sectionhard1}.
We will not repeat the calculations here.

\subsection{Double-magnon state}

The double-magnon state $|I_1I_2\rag$ and corresponding configuration probabilities in XXX chain are shown in appendix~\ref{appendixXXX}.

\subsubsection{Case I solution}

The total system Shannon entropy $H^{\I}_{00}(L)$ and subsystem Shannon entropy $H^{\I}_{00}(\ell)$ and mutual information $M^{\I}_{00}(\ell)$ in state $|00\rag$ are, respectively, the same as $H^\hard_{1^2}(L)$ (\ref{Hhard1hat2L}), $H^\hard_{1^2}(\ell)$ (\ref{Hhard1hat2ell}), and $M^\hard_{1^2}(\ell)$ (\ref{Ihard1hat2ell}).
We will not repeat the calculations here.

\subsubsection{Case II solution}

From these probabilities, we obtain for general parameters $L$, $\ell$, $I_1$ and $I_2$ the total system Shannon entropy and the subsystem Shannon entropy and mutual information
\bea
&& H^\II_{I_1I_2}(L) = - \sum_{1 \leq j_1 < j_2 \leq L} p^{\II}_{j_1j_2} \log p^{\II}_{j_1j_2},\label{HIII1I2L} \\
&& H^\II_{I_1I_2}(\ell) = - p^{A,\II}_0 \log p^{A,\II}_0
                        - \sum_{j=1}^\ell p^{A,\II}_j \log p^{A,\II}_j
                        - \sum_{1 \leq j_1 < j_2 \leq \ell} p^{A,\II}_{j_1j_2} \log p^{A,\II}_{j_1j_2},\label{HIII1I2ell} \\
&& M^\II_{I_1I_2}(\ell) = H^\II_{I_1I_2}(\ell) + H^\II_{I_1I_2} (L-\ell) -H^\II_{I_1I_2}(L),\label{IIII1I2ell}
\eea
where we have used the probabilities $p^\II_{j_1j_2}$ (\ref{XXXpIIj1j2}), $p^{A,\II}_0$ (\ref{pAII0}), $p^{A,\II}_j$ (\ref{pAIIj}) and $p^{A,\II}_{j_1j_2}$ (\ref{pAIIj1j2}).

\begin{figure}[t]
  \centering
  \includegraphics[height=0.28\textwidth]{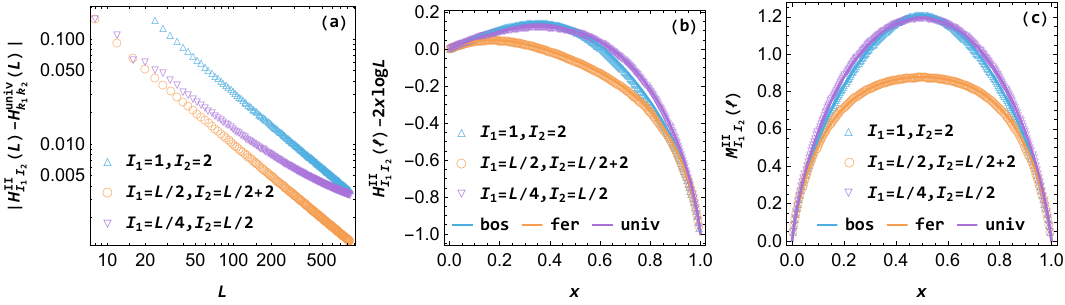}\\
  \caption{The total system Shannon entropy (left), subsystem Shannon entropy (middle) and mutual information (right) in the case II double-magnon state $|I_1I_2\rag$ of the XXX chain.
  The empty symbols are the exact numerical results.
  In the middle and right panels we have set $L=240$ for the exact results, and the solid lines are the corresponding analytical results in the scaling limit, namely the bosonic (bos) results $H^\bos_{k_1k_2}(\ell)$ (\ref{Hbosk1k2ellk12llL}) and $M^\bos_{k_1k_2}(\ell)$ (\ref{Ibosk1k2ellk12llL}) with $|k_{12}|=1$, fermionic (fer) results $H^\fer_{k_1k_2}(\ell)$ (\ref{Hferk1k2ellk12llL}) and $M^\fer_{k_1k_2}(\ell)$ (\ref{Iferk1k2ellk12llL}) with $|k_{12}|=1$, and the universal results $H^\univ_{k_1k_2}(\ell)$ (\ref{Hunivk1k2ell}) and $M^\univ_{k_1k_2}(\ell)$ (\ref{Iunivk1k2ell}).}
  \label{FigureIII1I2}
\end{figure}

We are interested in how the above results (\ref{HIII1I2L}), (\ref{HIII1I2ell}) and (\ref{IIII1I2ell}) behave in the scaling limit $L\to+\inf$, $\ell\to+\inf$ with fixed ratio $x={\ell}/{L}$.
We define the scaled Bethe numbers
\be
\io_1 = \lim_{L\to+\inf} \f{I_1}{L}, ~
\io_2 = \lim_{L\to+\inf} \f{I_2}{L}.
\ee
When $\io_1=\io_2=0$ or $\io_1=\io_2=1$ or $\io_1=0,\io_2=1$, we have $\th\to0$ and obtain
\bea
&&
\lim_{L\to+\inf} H^\II_{I_1I_2}(L) = H^\bos_{k_1k_2}(L), ~
\lim_{L\to+\inf} H^\II_{I_1I_2}(\ell) = H^\bos_{k_1k_2}(\ell), \nn\\
&&
\lim_{L\to+\inf} M^\II_{I_1I_2}(\ell) = M^\bos_{k_1k_2}(\ell), ~
   k_{12} = I_{12},
\eea
with $H^\bos_{k_1k_2}(L)$ (\ref{Hbosk1k2Lk12llL}), $H^\bos_{k_1k_2}(\ell)$ (\ref{Hbosk1k2ellk12llL}) and $M^\bos_{k_1k_2}(\ell)$ (\ref{Ibosk1k2ellk12llL}) being the bosonic results in the scaling limit.
When $\io_1=\io_2 \in (0,1)$, we have $\th\to\pi$ and the results approach to the fermionic results
\bea
&& \lim_{L\to+\inf} H^\II_{I_1I_2}(L) = H^\fer_{k_1k_2}(L), ~
   \lim_{L\to+\inf} H^\II_{I_1I_2}(\ell) = H^\fer_{k_1k_2}(\ell),  \nn\\
&& \lim_{L\to+\inf} M^\II_{I_1I_2}(\ell) = M^\fer_{k_1k_2}(\ell), ~
   k_{12} = I_{12} + 1,
\eea
with $H^\fer_{k_1k_2}(L)$ (\ref{Hferk1k2Lk12llL}), $H^\fer_{k_1k_2}(\ell)$ (\ref{Hferk1k2ellk12llL}) and $M^\fer_{k_1k_2}(\ell)$ (\ref{Iferk1k2ellk12llL}).
When $0 \leq \io_1 < \io_2 \leq 1$, excluding the case $\io_1=0,\io_2=1$, we have $\th\in[0,\pi)$ and the results approach to the universal results
\be
\lim_{L\to+\inf} H^\II_{I_1I_2}(L) = H^\univ_{k_1k_2}(L), ~
\lim_{L\to+\inf} H^\II_{I_1I_2}(\ell) = H^\univ_{k_1k_2}(\ell), ~
\lim_{L\to+\inf} M^\II_{I_1I_2}(\ell) = M^\univ_{k_1k_2}(\ell),
\ee
with $H^\univ_{k_1k_2}(L)$ (\ref{Hbosk1k2Lk12llL}), $H^\univ_{k_1k_2}(\ell)$ (\ref{Hunivk1k2ell}) and $M^\univ_{k_1k_2}(\ell)$ (\ref{Iunivk1k2ell}).
Note that the universal results do not depend on the actual values of the momenta $k_1,k_2$.
The exact numerical results of the total system Shannon entropy, subsystem Shannon entropy and mutual information, and the corresponding analytical results in the scaling limit are shown in Figure~\ref{FigureIII1I2}.

\subsubsection{Case IIIa solution}

From these probabilities, we obtain for general parameters $L$, $\ell$, and $I$ the total system Shannon entropy and the subsystem Shannon entropy and mutual information
\bea
&& H^\IIIa_{I_1I_2}(L) = - \sum_{1 \leq j_1 < j_2 \leq L} p^{\IIIa}_{j_1j_2} \log p^{\IIIa}_{j_1j_2}, \label{HIIIaI1I2L} \\
&& H^\IIIa_{I_1I_2}(\ell) = - p^{A,\IIIa}_0 \log p^{A,\IIIa}_0
                        - \sum_{j=1}^\ell p^{A,\IIIa}_j \log p^{A,\IIIa}_j
                        - \sum_{1 \leq j_1 < j_2 \leq \ell} p^{A,\IIIa}_{j_1j_2} \log p^{A,\IIIa}_{j_1j_2}, \label{HIIIaI1I2ell} \\
&& M^\IIIa_{I_1I_2}(\ell) = H^\IIIa_{I_1I_2}(\ell) + H^\IIIa_{I_1I_2} (L-\ell) -H^\IIIa_{I_1I_2}(L), \label{IIIIaI1I2ell}
\eea
with the probabilities $p^\IIIa_{j_1j_2}$ (\ref{XXXpIIIaj1j2}), $p^{A,\IIIa}_0$ (\ref{pAIIIa0}), $p^{A,\IIIa}_j$ (\ref{pAIIIaj}) and $p^{A,\IIIa}_{j_1j_2}$ (\ref{pAIIIaj1j2}).

For finite $v$ in the limit $L\to+\inf$, we have the tightly bound limit of the Shannon entropies and mutual information
\bea
&& H^\IIIa_{I_1I_2}(L) = \log L - \log(2\sinh v) + v \coth v, \label{HIIIaI1I2Ltight} \\
&& H^\IIIa_{I_1I_2}(\ell) = x [ \log L - \log(2\sinh v) + v \coth v ] - (1-x)\log(1-x), \label{HIIIaI1I2elltight} \\
&& M^\IIIa_{I_1I_2}(\ell) = - x \log x - (1-x)\log(1-x). \label{IIIIaI1I2elltight}
\eea
Note that the total system Shannon entropy (\ref{HIIIaI1I2Ltight}) has been obtained in \cite{Dong:2022fqg}.%
\footnote{In \cite{Dong:2022fqg} there is $q$ which is related to $v$ defined in this paper as $q=\ep^{-v}$.}
Although the bound state has finite width $1/v$, the mutual information (\ref{IIIIaI1I2elltight}) does not depend on $v$.
Further $v\to+\inf$ limit leads to the single-particle results
\bea
&& H^\IIIa_{I_1I_2}(L) = \log L, \\
&& H^\IIIa_{I_1I_2}(\ell) = x \log L- (1-x)\log(1-x), \\
&& M^\IIIa_{I_1I_2}(\ell) = - x \log x - (1-x)\log(1-x).
\eea

For $v=\f{u}{L}$ with finite $u$, which leads to
\be
u = - \lim_{L \to +\inf} L \log \Big| \cos \f{\pi I}{L} \Big|,
\ee
we get the loosely bound limit of the results
\bea
&& \hspace{-5mm} H^\IIIa_{I_1I_2}(L) = 2\log L+\log \Big( \f{\sinh u}{u} - 1 \Big) -2\log2
                    - \f{4}{\f{\sinh u}{u} - 1}
                      \int_0^{\f{u}{2}} \! \f{\dd y}{u} \sinh^2 y \log \sinh^2 y, \label{HIIIaI1I2Lloose} \\
&& \hspace{-5mm} H^\IIIa_{I_1I_2}(\ell) = 2 x \log L
                      + \Big( x + \f{\f{\sinh(u x)\sinh[u (1-x)]}{u^2} - x(1-x)}{\f{\sinh u}{u} - 1} \Big)
                        \log \Big( \f{\sinh u}{u} - 1 \Big) - 2x \log2 \nn\\
&& \hspace{-18mm} \phantom{H^\IIIa_{I_1I_2}(L) =}
                      - \Big( 1-x - \f{\f{\sinh(u x)\sinh[u (1-x)]}{u^2} - x(1-x)}{\f{\sinh u}{u} - 1} \Big)
                        \log \Big( 1-x - \f{\f{\sinh(u x)\sinh[u (1-x)]}{u^2} - x(1-x)}{\f{\sinh u}{u} - 1} \Big) \nn\\
&& \hspace{-18mm} \phantom{H^\IIIa_{I_1I_2}(L) =}
                      - \f{4}{\f{\sinh u}{u} - 1}
                        \int_0^{\f{x}{2}} \! \dd y \Big[ \f{\sinh[u (1-x)]\cosh(2uy)}{u} - (1-x) \Big] 
                                                   \log \Big[ \f{\sinh[u (1-x)]\cosh(2uy)}{u} - (1-x) \Big] \nn\\
&& \hspace{-18mm} \phantom{H^\IIIa_{I_1I_2}(L) =}
                      - \f{4}{\f{\sinh u}{u} - 1}
                        \int_0^x \! \dd y (x-y) \sinh^2 \Big[ u \Big(y-\f12\Big) \Big] \log \sinh^2 \Big[ u \Big(y-\f12\Big) \Big],\label{HIIIaI1I2ellloose} \\
&& \hspace{-5mm} M^\IIIa_{I_1I_2}(\ell) = H^\IIIa_{I_1I_2}(\ell) + H^\IIIa_{I_1I_2}(L-\ell) - H^\IIIa_{I_1I_2}(L). \label{IIIIaI1I2ellloose}
\eea

\begin{figure}[t]
  \centering
  \includegraphics[height=0.28\textwidth]{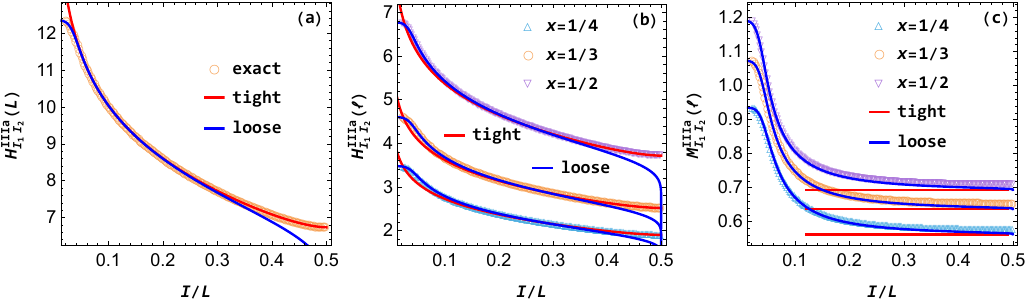}\\
  \caption{The total system Shannon entropy (left), subsystem Shannon entropy (middle) and mutual information (right) in the case IIIa double-magnon state $|I_1I_2\rag$ of the XXX chain.
  The empty symbols are the exact numerical results.
  The solid red and blue lines are, respectively, results of the tightly bound limit (\ref{HIIIaI1I2Ltight}), (\ref{HIIIaI1I2elltight}) and (\ref{IIIIaI1I2elltight}), and results of the loosely bound limit (\ref{HIIIaI1I2Lloose}), (\ref{HIIIaI1I2ellloose}) and (\ref{IIIIaI1I2ellloose}).
  We have set $L=840$.}
  \label{FigureIIIaI1I2}
\end{figure}

We show the total system Shannon entropy, subsystem Shannon entropy and mutual information in the case IIIa double-magnon state $|I_1I_2\rag$ of the XXX chain, as well as their tightly bound and loosely bound limits, in Figure~\ref{FigureIIIaI1I2}.
We see that in the range $I/L \in (0,0.5)$ the results of the loosely bound limit apply for relatively small $I/L\gtrsim0$ while the results of the tightly bound limit apply for relatively large $I/L\lesssim0.5$.
For the total system and subsystem Shannon entropies, there is the range $0.1 \lesssim I/L \lesssim 0.3$ in which both the results of the loosely bound limit and the results of the tightly bound limit apply.

\subsubsection{Case IIIb solution}

We obtain for general parameters $L$, $\ell$, and $I$ the total system Shannon entropy and the subsystem Shannon entropy and mutual information
\bea
&& H^\IIIb_{I_1I_2}(L) = - \sum_{1 \leq j_1 < j_2 \leq L} p^\IIIb_{j_1j_2} \log p^\IIIb_{j_1j_2}, \label{HIIIbI1I2L}\\
&& H^\IIIb_{I_1I_2}(\ell) = - p^{A,\IIIb}_0 \log p^{A,\IIIb}_0
                        - \sum_{j=1}^\ell p^{A,\IIIb}_j \log p^{A,\IIIb}_j
                        - \sum_{1 \leq j_1 < j_2 \leq \ell} p^{A,\IIIb}_{j_1j_2} \log p^{A,\IIIb}_{j_1j_2}, \label{HIIIbI1I2ell}\\
&& M^\IIIb_{I_1I_2}(\ell) = H^\IIIb_{I_1I_2}(\ell) + H^\IIIb_{I_1I_2} (L-\ell) -H^\IIIb_{I_1I_2}(L). \label{IIIIbI1I2ell}
\eea
with the probabilities $p^\IIIb_{j_1j_2}$ (\ref{XXXpIIIbj1j2}), $p^{A,\IIIb}_0$ (\ref{pAIIIb0}), $p^{A,\IIIb}_j$ (\ref{pAIIIbj}) and $p^{A,\IIIb}_{j_1j_2}$ (\ref{pAIIIbj1j2}).

For finite $v$, we obtain the same tightly bound results as those in case IIIa states
\bea
&& H^\IIIb_{I_1I_2}(L) = \log L - \log(2\sinh v) + v \coth v, \label{HIIIbI1I2Ltight}\\
&& H^\IIIb_{I_1I_2}(\ell) = x [ \log L - \log(2\sinh v) + v \coth v ] - (1-x)\log(1-x), \label{HIIIbI1I2elltight}\\
&& M^\IIIb_{I_1I_2}(\ell) = - x \log x - (1-x)\log(1-x). \label{IIIIbI1I2elltight}
\eea
Further $v\to+\inf$ limit leads to the single-particle results
\bea
&& H^\IIIb_{I_1I_2}(L) = \log L, \\
&& H^\IIIb_{I_1I_2}(\ell) = x \log L- (1-x)\log(1-x),\\
&& M^\IIIb_{I_1I_2}(\ell) = - x \log x - (1-x)\log(1-x).
\eea

For $v=\f{u}{L}$ with fixed $u$ in the scaling limit, we get the loosely bound limit of the results
\bea
&& \hspace{-5mm} H^\IIIb_{I_1I_2}(L) = 2\log L+\log \Big( \f{\sinh u}{u} + 1 \Big) -2\log2
                    - \f{4}{\f{\sinh u}{u} + 1}
                      \int_0^{\f{u}{2}} \! \f{\dd y}{u} \cosh^2 y \log \cosh^2 y,\label{HIIIbI1I2Lloose}\\
&& \hspace{-5mm} H^\IIIb_{I_1I_2}(\ell) = 2 x \log L
                      + \Big( x + \f{\f{\sinh(u x)\sinh[u (1-x)]}{u^2} + x(1-x)}{\f{\sinh u}{u} + 1} \Big)
                        \log \Big( \f{\sinh u}{u} + 1 \Big) - 2x \log2 \nn\\
&& \hspace{-18mm} \phantom{H^\IIIb_{I_1I_2}(L) =}
                      - \Big( 1-x - \f{\f{\sinh(u x)\sinh[u (1-x)]}{u^2} + x(1-x)}{\f{\sinh u}{u} + 1} \Big)
                        \log \Big( 1-x - \f{\f{\sinh(u x)\sinh[u (1-x)]}{u^2} + x(1-x)}{\f{\sinh u}{u} + 1} \Big) \nn\\
&& \hspace{-18mm} \phantom{H^\IIIb_{I_1I_2}(L) =}
                      - \f{4}{\f{\sinh u}{u} + 1}
                        \int_0^{\f{x}{2}} \! \dd y \Big[ \f{\sinh[u (1-x)]\cosh(2uy)}{u} + (1-x) \Big] 
                                                   \log \Big[ \f{\sinh[u (1-x)]\cosh(2uy)}{u} + (1-x) \Big] \nn\\
&& \hspace{-18mm} \phantom{H^\IIIb_{I_1I_2}(L) =}
                      - \f{4}{\f{\sinh u}{u} + 1}
                        \int_0^x \! \dd y (x-y) \cosh^2 \Big[ u \Big(y-\f12\Big) \Big] \log \cosh^2 \Big[ u \Big(y-\f12\Big) \Big], \label{HIIIbI1I2ellloose}
\eea\bea
&& \hspace{-5mm} M^\IIIb_{I_1I_2}(\ell) = H^\IIIb_{I_1I_2}(\ell) + H^\IIIb_{I_1I_2}(L-\ell) - H^\IIIb_{I_1I_2}(L).\label{IIIIbI1I2ellloose}
\eea
A further $u\to0$ limit of (\ref{HIIIbI1I2Lloose}), (\ref{HIIIbI1I2ellloose}), and (\ref{IIIIbI1I2ellloose}) leads to
\bea
&& H^\IIIb_{I_1I_2}(L) =  2\log L-\log 2, \label{eq1} \\
&& H^\IIIb_{I_1I_2}(\ell) = 2x \log L -2(1-x)\log(1-x) - x(2-x) \log2, \label{eq2} \\
&& M^\IIIb_{I_1I_2}(\ell) = -x^2\log x^2- 2x(1-x)\log [2x(1-x)]-(1-x)^2\log (1-x)^2, \label{eq3}
\eea
which are the same as the results of two identical classical particles $H^\cl_{1^2}(L)$ (\ref{Hcl1hat2L}), $H^\cl_{1^2}(\ell)$ (\ref{Hcl1hat2ell}), and $M^\cl_{1^2}(\ell)$ (\ref{Icl1hat2ell}).
For $v=\f{w}{L^2}$ with fixed $w$ in the scaling limit, we get the same results (\ref{eq1}), (\ref{eq2}), and (\ref{eq3}).
In the limit where $v$ approaches zero, known as the loosely bound limit, we can observe from equation (\ref{XXXIIIbp1p2}) that $p_1$ becomes equal to $p_2$ and $\theta$ becomes zero. This implies that the two magnons exhibit behavior similar to two bosonic particles with identical momentum, which corresponds to the scenario analyzed in subsection~\ref{subsectiondpsk2}. Consequently, this elucidates why the outcomes align with those obtained for two identical classical particles.

\begin{figure}[t]
  \centering
  \includegraphics[height=0.28\textwidth]{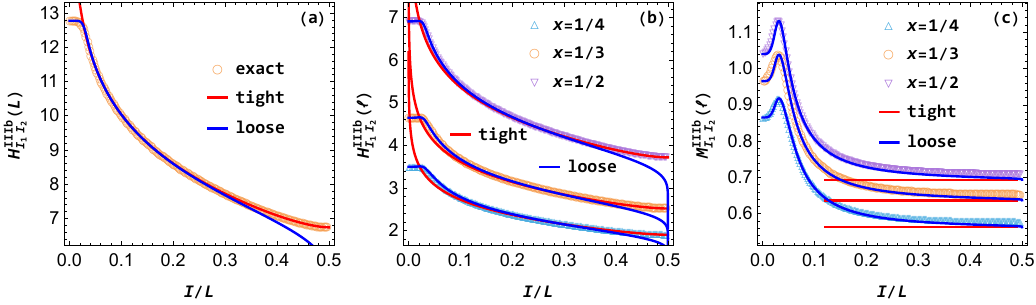}\\
  \caption{The total system Shannon entropy (left), subsystem Shannon entropy (middle) and mutual information (right) in the case IIIb double-magnon state $|I_1I_2\rag$ of the XXX chain.
  The empty symbols are the exact numerical results.
  The solid red and blue lines are, respectively, the tightly bound results (\ref{HIIIbI1I2Ltight}), (\ref{HIIIbI1I2elltight}) and (\ref{IIIIbI1I2elltight}), and the loosely bound results (\ref{HIIIbI1I2Lloose}), (\ref{HIIIbI1I2ellloose}) and (\ref{IIIIbI1I2ellloose}).
  We have set $L=840$.}
  \label{FigureIIIbI1I2}
\end{figure}

We show the total system Shannon entropy, subsystem Shannon entropy and mutual information in the case IIIb double-magnon state $|I_1I_2\rag$ of the XXX chain, as well as their tightly and loosely bound limits, in Figure~\ref{FigureIIIbI1I2}.

\section{States with three and four quasiparticles} \label{sectionthreenfour}

To verify the properties discovered for the double-particle states in the bosonic, fermionic, and XXX chains, in this section we calculate numerically the Shannon entropy and mutual information in states with three and four quasiparticles.

In the free bosonic chain we consider the triple-particle states $|k_1^2k_2\rag$ and $|k_1k_2k_3\rag$ and the quadruple-particle states $|k_1^3k_2\rag$, $|k_1^2k_2^2\rag$, $|k_1^2k_2k_3\rag$ and $|k_1k_2k_3k_4\rag$. In the free fermionic chain, we consider the triple-particle state $|k_1k_2k_3\rag$ and the quadruple-particle state $|k_1k_2k_3k_4\rag$.
We set $k_i=1+(i-1)\d k$ with $i=1,2,3,4$ and $\d k =1,2,\cdots$.

In the XXX chain, we consider the triple-magnon state $|\cI_1\rag$ with Bethe quantum numbers $\cI_1=(1,1+\d I,1+2\d I)$ and $\d I=1,3,5$, which corresponds to the bosonic triple-particle state $|K_1\rag$ with $K_1=(1,1+\d k,1+2\d k)$ and $\d k=\d I$ in the scaling limit.
We also consider the triple-magnon state $|\cI_2\rag$  with Bethe quantum numbers $\cI_2=(\f{L}{2},\f{L}{2}+\d I,\f{L}{2}+2\d I)$ and $\d I=2,4,6$, which corresponds to the fermionic triple-particle state $|K_2\rag$ with $K_2=(1,1+\d k,1+2\d k)$ and $\d k=\d I-1$.
Note the difference that in the bosonic chain $\d k=\d I$ and in the fermionic chain $\d k=\d I-1$.

\begin{figure}[t]
  \centering
  \includegraphics[height=0.28\textwidth]{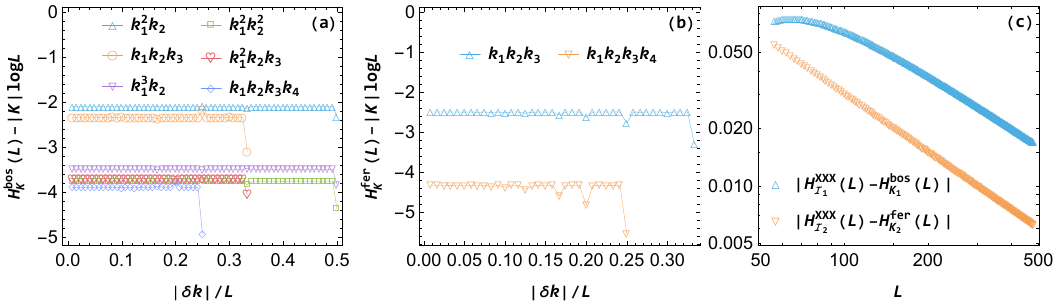}\\
  \caption{In the left and middle panels, we show the total system Shannon entropy in the triple-particle and quadruple-particle states of the free bosonic and fermionic chains, respectively. We have set $L=120$ and $k_i=1+(i-1)\d k$ with $i=1,2,3,4$.
  In the right panel, we show the difference of total system Shannon entropy in the XXX chain and the corresponding result in the free bosonic/fermionic chain.
  For the state $\cI_1=(1,2,3)$, there is the corresponding bosonic state $K_1=(1,2,3)$, and for the state $\cI_2=(\f{L}{2},\f{L}{2}+2,\f{L}{2}+4)$, there is the corresponding fermionic state $K_2=(1,2,3)$.}
  \label{FiguretotSE}
\end{figure}

We present numerical results of the total system Shannon entropy in these triple- and quadruple-particle states in the left and middle panels of Figure~\ref{FiguretotSE}. For each state, the total system Shannon entropy approaches a universal value for almost all values of $\d k$, except for some exceptional cases when $\d k/L$ has a finite numerator and denominator in the scaling limit. In both the bosonic and fermionic chains, the universal values for the states $|k_1k_2k_3\rag$ and $|k_1k_2k_3k_4\rag$ are different, contrasting with the double-particle state $|k_1k_2\rag$ which has a universal quantum value. In the right panel of Figure~\ref{FiguretotSE}, we compare the total system Shannon entropy in the XXX chain and the free bosonic/fermionic chain. We observe that in the scaling limit, the results for certain states in the XXX chain approach the corresponding values in the free bosonic/fermionic chain.

\begin{figure}[tp]
  \centering
  \includegraphics[height=0.56\textwidth]{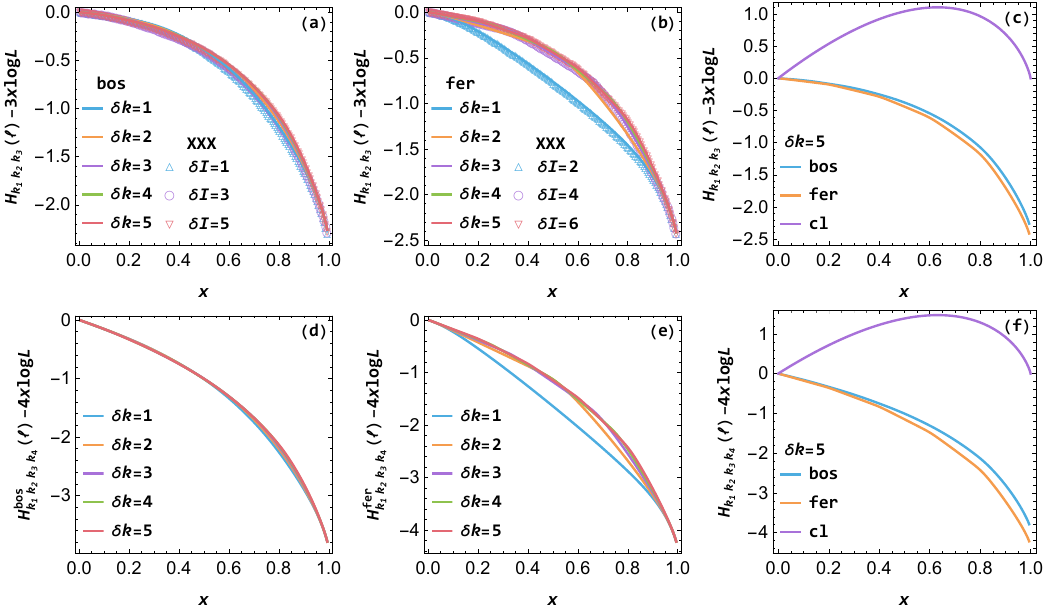}\\
  \caption{The subsystem Shannon entropy in triple-particle states (top panels) and quadruple-particle states (bottom panels) of the free bosonic and fermionic chains and the XXX chain.
  The length of the chain is $L=180$.}
  \label{FiguresubSE}
\end{figure}

\begin{figure}[tp]
  \centering
  \includegraphics[height=0.56\textwidth]{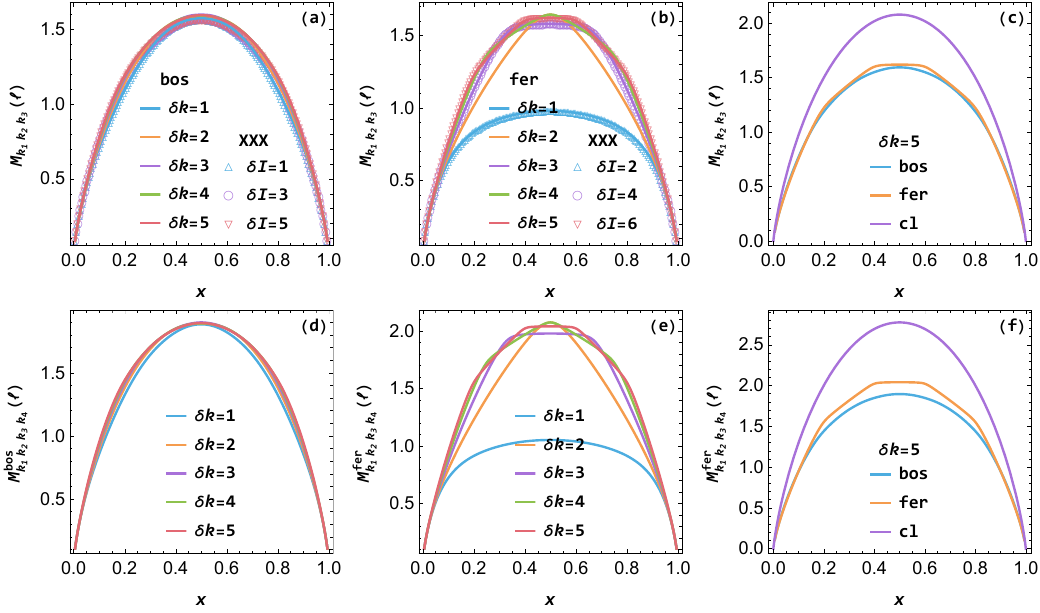}\\
  \caption{The subsystem Shannon mutual information in triple-particle states (top panels) and quadruple-particle states (bottom panels) of the free bosonic and fermionic chains and the XXX chain.
  We have used $L=180$.}
  \label{FigureMI}
\end{figure}

We show results of the subsystem Shannon entropy and mutual information in, respectively, figures~\ref{FiguresubSE} and \ref{FigureMI}.
With the increase of the momentum difference $\d k$, the subsystem Shannon entropy and mutual information approach some universal bosonic and fermionic values, as shown in panels (a), (b), (d) and (e) of figures~\ref{FiguresubSE} and panels (a), (b), (d) and (e) of figures~\ref{FigureMI}, but the bosonic and fermionic values are generally distinct and also different from the results of classical particles, as shown in panels (c) and (f) of figures~\ref{FiguresubSE} and panels (c) and (f) of figures~\ref{FigureMI}.
In panels (a) and (b) of figures~\ref{FiguresubSE} and panels (a) and (b) of figures~\ref{FigureMI}, we also show that results of certain states in the XXX chain approach the corresponding free bosonic/ferminic results.

The main conclusion of the paper is that there does not exist a classical description for the local basis Shannon entropy in quantum chains, as exemplified in the previous sections and this section.
Additionally, it has been shown that contributions to the Shannon entropy from quasiparticles with large momentum differences do not separate.

We denote a general quasiparticle excited state as $|K\rag$ with $K$ being the set of the momenta of the excited quasiparticles. The results for double-particle states in the previous sections and the results for triple- and quadruple-particle states in this section support the general form of the subsystem Shannon entropy in the scaling limit
\be \label{HKell}
H_K(\ell) = |K| x \log L + \d H_K(x),
\ee
where $|K|$ is the total number of excited quasiparticles and $\d H_K(x)$ is a function of the ratio $x=\ell/L$.
On the right hand side of (\ref{HKell}), the universal logarithmic divergent part is the same as that in the classical results (\ref{Icl1hatrell}) and (\ref{Hcl1r12r2cdotssr2ell}), and this guarantees a finite subsystem Shannon mutual information in the scaling limit.
For double-particle states, we have observed a universal quantum result in the limit $1\ll |k_{12}| \ll L$
\be
\d H_{k_1k_2}^\bos(x)=\d H_{k_1k_2}^\fer(x),
\ee
which is, however, generally {\it not} true for general triple- and quadruple-particle states.
In the limit $1\ll |k_{12}| \ll L$, there is the Shannon mutual information
\be
M_{k_1k_2}^\bos(x)=M_{k_1k_2}^\fer(x),
\ee
which is also {\it not} true for general multi-particle states.
Generally, in large momentum difference limit, there exist distinct universal bosonic and fermionic results, which are also different from the classical results.

\section{Conclusion and discussion} \label{sectionConclusion}

In this paper we have presented an analysis of the total system Shannon entropy, subsystem Shannon entropy, and subsystem Shannon mutual information in quasiparticle excited states of free bosonic and fermionic chains, as well as the XXX chain.
We compared the results across different spin chains and contrasted them with those of classical particles.
Our findings indicate that, like the entanglement entropy, the formulas for the Shannon entropy in free bosonic and fermionic chains also apply to the Shannon entropy in the XXX chain, subject to certain limits.
Additionally, we found that even when quasiparticles have large momentum differences, their contributions to the Shannon entropy do not decouple, unlike the entanglement entropy.
In addition, we have discovered distinct universal bosonic and fermionic formulas for the total system Shannon entropy, subsystem Shannon entropy, and mutual information in the limit of large momentum differences, which are also distinct from any classical particle-based results.
In other words, we did not observe any semiclassical quasiparticle picture that could reproduce the results of the Shannon entropy in quantum spin chains.

\begin{figure}[t]
  \centering
  \includegraphics[width=0.75\textwidth]{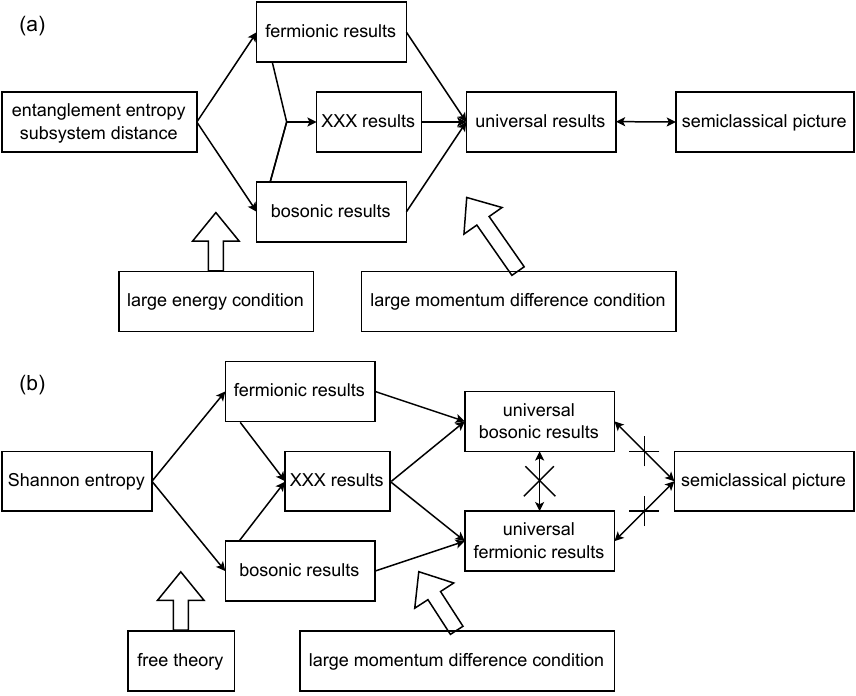}\\
  \caption{Summary of the different pictures for (a) the entanglement entropy and subsystem distance and (b) the Shannon entropy in the local particle number basis.}
  \label{FigureEntanglementDistanceShannon}
\end{figure}

In Figure~\ref{FigureEntanglementDistanceShannon}, we present the different results for the entanglement entropy, subsystem distance, and the local particle number basis Shannon entropy. For simplicity, we only calculated the Shannon entropy in free theories for fermionic and bosonic chains in this paper, which are a special case of the large energy condition.
Unlike the entanglement entropy and subsystem distance, the Shannon entropy contributions from quasiparticles with large momentum differences do not decouple. As a result, the combined fermionic and bosonic results are generally not applicable to the XXX chain. However, the fermionic and bosonic results can still be applied separately to the XXX chain within certain limits.
Another notable distinction is that the semiclassical picture holds true for the universal entanglement entropy and subsystem distance, but it fails to explain the universal local particle number basis Shannon entropy.

In the main text of the paper, we have calculated the Shannon entropy using a particular basis of local states. It is essential to note that the definition of the Shannon entropy is dependent on the chosen basis. If one selects the eigenstates of the density matrix as the basis, the Shannon entropy will become equivalent to the von Neumann entropy for both pure and mixed quantum states of the total system or its subsystem. For quantum spin chains, the eigenstates of the subsystem particle number operator are also a natural basis.
In Appendix~\ref{appendixPNPD}, we have studied the Shannon entropy of the subsystem particle number probability distribution in free bosonic and fermionic chains. We have found that the semiclassical quasiparticle picture applies in the large momentum difference limit. In Appendix~\ref{appendixXB}, we have calculated the Shannon entropy in the local basis of the $\s_j^x$ eigenstates in the XXX chain. The results are significantly different from those obtained using the $\s_j^z$ basis. Even in the single-magnon state, the $\s_j^x$ basis Shannon entropy is already nontrivial. In certain limits, there exist forms of universal results for which no semiclassical quasiparticle picture applies.

We have found Shannon entropy in local bases for which no semiclassical quasiparticle picture applies.
At the same time, there are the entanglement entropy, subsystem distance and Shannon entropy of the subsystem particle number probability distribution for which the semiclassical quasiparticle picture applies in the large momentum difference limit.
We anticipate that the semiclassical picture applies to coarse-grained properties but not to fine-grained properties.
In other words, if one scrutinizes the microscopic details of a quasiparticle excited state, one can distinguish between quasiparticles and real particles. However, for a measure of certain macroscopic properties, quasiparticles behave similarly to real particles.

In our analysis, we have focused on models that encompass a single species of quasiparticles. In appendix \ref{appendixSSH}, we provide a concise discussion on the entanglement entropy and Shannon entropy within the context of the SSH model \cite{Su:1979ua,Su:1979wc}, which incorporates two independent sets of fermionic modes. Our findings indicate that the individual contributions of these distinct fermionic species to both entanglement entropy and Shannon entropy are independent of each other. It would be of interest to investigate the behavior of Shannon entropy in alternative models that feature multiple species of quasiparticles.
In this paper we have focused on translation invariant states in periodic chains. It would be intertwisting to consider states of inhomogeneous models such as open chains.

Experimental realization of excited states in spin chains was demonstrated in trapped atomic ion systems \cite{Jurcevic:2014qfa,Jurcevic:2015inc,Zhang:2017wrv}. Recently, a proposal has been put forward to utilize neutral atom arrays to simulate fermionic many-body systems \cite{Gonzalez-Cuadra:2023rex}. It would be fascinating to measure the Shannon entropy in quasiparticle excited states of spin chains and compare the experimental outcomes with the results presented in this paper.
Within this context, measuring the Shannon entropy in quasiparticle excited states of spin chains offers a window into their quantum correlations. The Shannon entropy, as a quantitative measure of information content or uncertainty in a probability distribution, can provide insights into the distribution of quantum states in a system and how they evolve under various interactions and perturbations. Comparing the experimental outcomes with the theoretical predictions presented in this paper (or similar theoretical frameworks) would not only validate the accuracy of the experimental platform but also shed light on the microscopic mechanisms governing the quasiparticle dynamics. Furthermore, such experiments could potentially uncover novel quantum phenomena, such as unconventional quasiparticle statistics or emergent interactions.

\section*{Acknowledgements}

We thank M.~A.~Rajabpour for reviewing an earlier version of the draft and providing valuable feedback and suggestions.
JZ acknowledges the support from the National Natural Science Foundation of China (NSFC) through grant number 12205217.

\appendix

\section{States and probabilities in free bosonic chain}\label{appendixBoson}

In this appendix, we show calculation details for the free bosonic chain.
We mainly show the constructions of the single-particle and double-particle excited states and the corresponding probabilities of subsystem configurations.

\subsection{Quasiparticle excited states}

The free bosonic chain has the Hamiltonian
\be
H = \sum_{j=1}^L \Big( a_j^\dag a_j + \f12 \Big).
\ee
We define the global modes
\be
b_k^\dag \equiv \f{1}{\sqrt{L}} \sum_{j=1}^L \ep^{\f{2\pi\ii j k}{L}} a_j^\dag, ~
b_k \equiv \f{1}{\sqrt{L}} \sum_{j=1}^L \ep^{-\f{2\pi\ii j k}{L}} a_j, ~
k=0,1,\cdots,L-1.
\ee
The ground state is defined as
\be
a_j |G\rag = b_k |G\rag = 0, ~ \forall j, \forall k.
\ee
Using the global modes $b_k^\dag$, one could construct the general translation-invariant global state
\be
|k_1^{r_1}k_2^{r_2}\cdots k_s^{r_s}\rag = \f{( b^\dag_{k_1} )^{r_1} ( b^\dag_{k_2} )^{r_2} \cdots ( b^\dag_{k_s} )^{r_s}}{\sr{r_1!r_2!\cdots r_s!}} | G \rag.
\ee

In the free bosonic chain, the natural local basis at each site is the eigenstates of the operator of the excitation number.
One could use the local modes $a_j^\dag$ to construct the general locally excited state
\be
|j_1^{r_1}j_2^{r_2}\cdots j_s^{r_s}\rag = \f{( a^\dag_{j_1} )^{r_1} ( a^\dag_{j_2} )^{r_2} \cdots ( a^\dag_{j_s} )^{r_s}}{\sr{r_1!r_2!\cdots r_s!}} | G \rag,
\ee
which we will call local state for short.
In the Hilbert space of the subsystem $A=[1,\ell]$, one has similarly the subsystem ground state $|G\rag_A$ defined as
\be
a_j |G\rag_A = 0, ~ \forall j \in A,
\ee
as well as the subsystem local excited states
\be
|j_1^{r_1}j_2^{r_2}\cdots j_s^{r_s}\rag_A
=
\f{( a^\dag_{j_1} )^{r_1} ( a^\dag_{j_2} )^{r_2} \cdots ( a^\dag_{j_s} )^{r_s}}{\sr{r_1!r_2!\cdots r_s!}} | G \rag_A, ~
j_1,j_2,\cdots,j_s \in A.
\ee

\subsection{Single-particle state $|k\rag$}\label{appendixkbos}

We first consider the global single-particle state $|k\rag = b_k^\dag |G\rag$, which could be written in terms of local states as
\be
|k\rag = \f{1}{\sqrt{L}} \sum_{j=1}^L \ep^{\f{2\pi\ii j k}{L}} |j\rag.
\ee
In state $|k\rag$, there are $L$ possible local states $|j\rag$ with $j \in [1,L]$, and the corresponding probabilities are
\be \label{bosspspj}
p_j = \f{1}{L}, ~ j \in [1,L].
\ee

For the subsystem $A$, there are $\ell+1$ possible local states and the corresponding probabilities are listed in Table~\ref{tablebosksub}.
The probabilities of the subsystem local states in Table~\ref{tablebosksub} could be calculated either as the diagonal entries of the reduced density matrix in the local basis or as marginal probabilities of the probability distribution (\ref{bosspspj}) of the whole system.
In this paper we will adopt the latter approach.
Explicitly, we have used
\bea
&& p_{A,0} = \sum_{j=\ell+1}^L p_j,\nn\\
&& p_{A,j} = p_j, ~ j \in [1,\ell].
\eea

\begin{table}[h]
  \centering
\begin{tabular}{|c|c|c|c|} \hline
   local states & probabilities        & ranges           & numbers \\ \hline
   $|G\rag_A$   & $p_{A,0} = 1 - x$    & -                & 1       \\ \hline
   $|j\rag_A$   & $p_{A,j} = \f{1}{L}$ & $j \in [1,\ell]$ & $\ell$  \\ \hline
\end{tabular}
  \caption{The local states and probabilities of the subsystem $A$ in the single particle state $|k\rag$ of the free bosonic chain. We have used $x=\f{\ell}{L}$.}
  \label{tablebosksub}
\end{table}

\subsection{Double-particle state $|k^2\rag$}

In terms of the local states, the global double-particle state $|k^2\rag=\f{1}{\sr{2}}(b_k^\dag)^2|G\rag$ could be written as
\be
|k^2\rangle = \frac {1}{L}\sum_{j=1}^L \ep^\frac{4\pi ijk}{L} |j^2 \rangle
                                            +\frac{\sqrt2}{L}\sum_{1\le j_1<j_2\le L}e^\frac{2\pi i k(j_1+j_2)}{L}|j_1j_2\rangle.
\ee
There are $\f{L(L+1)}{2}$ possible local states, and the probabilities are shown in Table~\ref{tableboskhat2tot}, which are the same as the probabilities of the configuration of two identical classical soft-core particles in Table~\ref{tablesoft2indtot}.

\begin{table}[h]
  \centering
\begin{tabular}{|c|c|c|c|} \hline
   local states  & probabilities             & ranges                    & numbers         \\ \hline
   $|j^2\rag$    & $p_{j^2} = \f{1}{L^2}$    & $j \in [1,L]$             & $L$             \\ \hline
   $|j_1j_2\rag$ & $p_{j_1j_2} = \f{2}{L^2}$ & $1 \leq j_1 < j_2 \leq L$ & $\f{L(L-1)}{2}$ \\ \hline
\end{tabular}
  \caption{The local states and probabilities of the total system in the double-particle state $|k^2\rag$ of the free bosonic chain.}
  \label{tableboskhat2tot}
\end{table}

For the subsystem $A$, there are $\f{(\ell+1)(\ell+2)}{2}$ possible local states and the corresponding probabilities are shown in Table~\ref{tableboskhat2sub}.
The probabilities in Table~\ref{tableboskhat2sub} are actually the marginal probabilities of the probability distribution in Table~\ref{tableboskhat2tot}.

\begin{table}[h]
  \centering
\begin{tabular}{|c|c|c|c|} \hline
   local states    & probabilities             & ranges                       & numbers \\ \hline
   $|G\rag_A$      & $p_{A,0} = (1 - x)^2$     & -                            & 1       \\ \hline
   $|j\rag_A$      & $p_{A,j} = \f{2(1-x)}{L}$ & $j \in [1,\ell]$             & $\ell$  \\ \hline
   $|j^2\rag_A$    & $p_{A,j^2} = \f{1}{L^2}$  & $j \in [1,\ell]$             & $\ell$     \\ \hline
   $|j_1j_2\rag_A$ & $p_{A,j_1j_2}=\f{2}{L^2}$ & $1 \leq j_1 < j_2 \leq \ell$ & $\f{\ell(\ell-1)}{2}$ \\ \hline
\end{tabular}
  \caption{The local states and probabilities of the subsystem $A$ in the double-particle state $|k^2\rag$ of the free bosonic chain.}
  \label{tableboskhat2sub}
\end{table}

\subsection{Double-particle state $|k_1k_2\rag$}

The double-particle state $|k_1k_2\rag$ could be written in terms of local states as
\be
|k_1k_2\rag = \frac{\sqrt2}{L} \sum_{j=1}^{L} e^\frac{2\pi \ii j(k_1+k_2)}{L}|j^2\rangle
            + \frac{2}{L} \sum_{1\le j_1<j_2\le L}e^{\frac{\pi\ii}{L}(j_1+j_2)(k_1+k_2)}\cos\frac{\pi j_{12}k_{12}}{L}|j_1j_2\rangle,
\ee
with the shorthand $j_{12} \equiv j_1-j_2$ and $k_{12} \equiv k_1-k_2$.
As there is period $L$ for the momentum $k \cong k+L$, we only need to consider the case with $1\leq |k_{12}| \leq \f{L}{2}$.
There are $\f{L(L+1)}{2}$ possible local states, and the probabilities are shown in Table~\ref{tablebosk1k2tot}.

\begin{table}[h]
  \centering
\begin{tabular}{|c|c|c|c|} \hline
   local states  & probabilities             & ranges                    & numbers         \\ \hline
   $|j^2\rag$    & $p_{j^2} = \f{2}{L^2}$        & $j \in [1,L]$             & $L$             \\ \hline
   $|j_1j_2\rag$ & $p_{j_1j_2} = \f{4}{L^2}\cos^2\f{\pi j_{12} k_{12}}{L}$ & $1 \leq j_1 < j_2 \leq L$ & $\f{L(L-1)}{2}$ \\ \hline
\end{tabular}
  \caption{The local states and probabilities of the total system in the double-particle state $|k_1k_2\rag$ of the free bosonic chain.}
  \label{tablebosk1k2tot}
\end{table}

For the subsystem $A$, there are $\f{(\ell+1)(\ell+2)}{2}$ possible local states and the probabilities are shown in Table~\ref{tablebosk1k2sub}, where we have the marginal probabilities
\bea
&& p_{A,0} = \sum_{j=\ell+1}^L p_{j^2} + \sum_{\ell+1\leq j_1 < j_2 \leq L} p_{j_1j_2} = (1 - x)^2+\f{\sin^2(\pi k_{12} x)}{L^2\sin^2\f{\pi k_{12}}{L}}, \label{bosonp0} \\
&& p_{A,j} = \sum_{j_2=\ell+1}^L p_{jj_2} = \f{2(1-x)}{L} - \f{2\sin(\pi k_{12} x)\cos\f{2\pi k_{12} (j-\f{\ell+1}{2})}{L}}{L^2\sin\f{\pi k_{12}}{L}}, \label{bosonpj}
\eea
with $p_{j^2}$ and $p_{j_1j_2}$ being defined in Table~\ref{tablebosk1k2tot}.

\begin{table}[h]
  \centering
\begin{tabular}{|c|c|c|c|} \hline
   local states    & probabilities           & ranges                        & numbers \\ \hline
   $|G\rag_A$      & $p_{A,0}$ (\ref{bosonp0})   & -                             & 1       \\ \hline
   $|j\rag_A$      & $p_{A,j}$ (\ref{bosonpj})   & $j \in [1,\ell]$              & $\ell$  \\ \hline
   $|j^2\rag_A$    & $p_{A,j^2} = \f{2}{L^2}$  & $j \in [1,\ell]$              & $\ell$  \\ \hline
   $|j_1j_2\rag_A$ & $p_{A,j_1j_2} = \f{4}{L^2}\cos^2\f{\pi j_{12} k_{12}}{L}$ & $1 \leq j_1 < j_2 \leq \ell$ & $\f{\ell(\ell-1)}{2}$ \\ \hline
\end{tabular}
  \caption{The local states and probabilities of the subsystem $A$ in the double-particle state $|k_1k_2\rag$ of the free bosonic chain.}
  \label{tablebosk1k2sub}
\end{table}

\section{States and probabilities in free fermionic chain}\label{appendixFermion}

In this appendix, we show calculation details for the free fermionic chain.

\subsection{Quasiparticle excited states}

The free fermionic chain has the Hamiltonian
\be
H = \sum_{j=1}^L \Big( a_j^\dag a_j - \f12 \Big).
\ee
We define the global modes
\be
b_k^\dag \equiv \f{1}{\sqrt{L}} \sum_{j=1}^L \ep^{\f{2\pi\ii j k}{L}} a_j^\dag, ~
b_k \equiv \f{1}{\sqrt{L}} \sum_{j=1}^L \ep^{-\f{2\pi\ii j k}{L}} a_j, ~
k=0,1,\cdots,L-1.
\ee
The ground state is defined as
\be
a_j |G\rag = b_k |G\rag = 0, ~ \forall j, \forall k.
\ee
Using the global modes $b_k^\dag$, one could construct the general global state
\be
|k_1k_2\cdots k_s\rag = b_{k_1}^\dag b_{k_2}^\dag \cdots b_{k_s}^\dag |G\rag.
\ee

Like the free bosonic chain, the natural local basis at each site is the eigenstates of the excitation number operator.
One could also use the modes $a_j^\dag$ to construct the general locally excited state
\be
|j_1j_2\cdots j_s\rag = a_{j_1}^\dag a_{j_2}^\dag \cdots a_{j_s}^\dag |G\rag.
\ee
In the Hilbert space of the subsystem $A=[1,\ell]$, one has similarly the subsystem ground state $|G\rag_A$ defined as
\be
a_j |G\rag_A = 0, ~ \forall j \in A,
\ee
as well as the subsystem local states
\be
|j_1j_2\cdots j_s\rag_A = a_{j_1}^\dag a_{j_2}^\dag \cdots a_{j_s}^\dag |G\rag_A, ~
j_1,j_2,\cdots,j_s \in A.
\ee

\subsection{Single-particle state $|k\rag$}

The results for the single-particle state $|k\rag$ in the free fermionic chain are the same as those in the free bosonic chain in appendix~\ref{appendixkbos}. We will not repeat it here.

\subsection{Double-particle state $|k_1k_2\rag$}

We write the double-particle state $|k_1k_2\rag$ in terms of local states as
\be
|k_1k_2\rag = \frac{2\ii}{L} \sum_{1\le j_1<j_2\le L}e^{\frac{\pi\ii}{L}(j_1+j_2)(k_1+k_2)}\sin\frac{\pi j_{12}k_{12}}{L}|j_1j_2\rangle,
\ee
with $j_{12} = j_1-j_2$ and $k_{12} = k_1-k_2$.
There are $\f{L(L-1)}{2}$ possible local states $|j_1j_2\rag$ with $1 \leq j_1 < j_2 \leq L$, and the corresponding probabilities are
\be \label{ferdpspj1j2}
p_{j_1j_2} = \f{4}{L^2}\sin^2\f{\pi j_{12} k_{12}}{L}, ~ 1 \leq j_1 < j_2 \leq L.
\ee

For the subsystem $A$, there are $\f{\ell(\ell+1)}{2}+1$ possible local states and the probabilities are shown in Table~\ref{tableferk1k2sub}, where we have the marginal probabilities
\bea
&& p_{A,0} = \sum_{\ell+1\leq j_1 < j_2 \leq L} p_{j_1j_2}
           = (1 - x)^2 - \f{\sin^2(\pi k_{12} x)}{L^2\sin^2\f{\pi k_{12}}{L}},\label{fermionp0} \\
&& p_{A,j} = \sum_{j_2=\ell+1}^L p_{jj_2}
           = \f{2(1-x)}{L} + \f{2\sin(\pi k_{12} x)\cos\f{2\pi k_{12} (j-\f{\ell+1}{2})}{L}}{L^2\sin\f{\pi k_{12}}{L}},\label{fermionpj}
\eea
with $p_{j_1j_2}$ being defined in (\ref{ferdpspj1j2}).

\begin{table}[h]
  \centering
\begin{tabular}{|c|c|c|c|} \hline
   local states    & probabilities           & ranges                        & numbers \\ \hline
   $|G\rag_A$      & $p_{A,0}$ (\ref{fermionp0})   & -                             & 1       \\ \hline
   $|j\rag_A$      & $p_{A,j}$ (\ref{fermionpj})   & $j \in [1,\ell]$              & $\ell$  \\ \hline
   $|j_1j_2\rag_A$ & $p_{A,j_1j_2} = \f{4}{L^2}\sin^2\f{\pi j_{12} k_{12}}{L}$ & $1 \leq j_1 < j_2 \leq \ell$ & $\f{\ell(\ell-1)}{2}$ \\ \hline
\end{tabular}
  \caption{The local states and probabilities of the subsystem $A$ in the double-particle state $|k_1k_2\rag$ of the free fermionic chain.}
  \label{tableferk1k2sub}
\end{table}

\section{States and probabilities in XXX chain}\label{appendixXXX}

In this appendix, we show calculation details for the spin-1/2 XXX chain.

\subsection{Magnon excited states}

The spin-1/2 XXX chain has the Hamiltonian
\be \label{XXXHamiltonian}
H = -\f14 \sum_{j=1}^L ( \s_j^x\s_{j+1}^x + \s_j^y\s_{j+1}^y + \s_j^z\s_{j+1}^z ) - \f{h}{2} \sum_{j=1}^L \s_j^z,
\ee
with a positive transverse field $h>0$ and periodic boundary conditions for the Pauli matrices $\s_{L+1}^{x,y,z}=\s_1^{x,y,z}$.
For simplicity, we also require that the number of sites $L$ is four times of an integer.
The unique ground state is
\be
|G\rag = |\!\uparrow\uparrow\cdots\uparrow\rag.
\ee
The low-lying eigenstates are magnon excited states in the ferromagnetic phase and can be obtained from the coordinate Bethe ansatz \cite{Caux:2014uuq,Karbach:1998abi}.
One may use the Bethe numbers of the excited magnons $\{I_1,I_2,\cdots,I_s\}$ to denote magnon excited states as $|I_1 I_2 \cdots I_s\rag$.

The presence of the transverse field in the Hamiltonian (\ref{XXXHamiltonian}) makes the eigenstates of $\s_j^z$ to be a natural local basis at each site for the XXX chain.%
\footnote{In appendix~\ref{appendixXB}, we show the Shannon entropy in the local basis of $\s_j^x$ eigenstates, and the results are very different from the those in $\s_j^z$ basis.}
There are also local states, such as
\be
|j\rag = |\!\cdots\downarrow_j\cdots\rag,
\ee
in which only the site $j$ has downward spin and all other sites have upward spins,
and
\be
|j_1j_2\rag = |\!\cdots\downarrow_{j_1}\cdots\downarrow_{j_2}\cdots\rag,
\ee
in which only the sites $j_1,j_2$ have downward spins and all other sites have upward spins.
For the subsystem $A$, there are the subsystem local states $|G\rag_A$ in which all the sites in $A$ has upward spins,
$|j\rag_A$ in which only the site $j$ in $A$ has downward spin,
and $|j_1j_2\rag_A$ in which only the sites $j_1,j_2$ in $A$ have downward spins.

\subsection{Single-magnon state}

The single magnon state takes the form
\be
|I\rag = \f{1}{\sqrt{L}} \sum_{j=1}^L \ep^{\f{2\pi\ii j I}{L}} |j\rag,
\ee
with the Bethe number $I=0,1,\cdots,L-1$.
The probability distributions are the same as those in the free bosonic chain in appendix~\ref{appendixkbos}.

\subsection{Double-magnon state}

We consider the double-magnon state
\be
|I_1I_2\rag = \f{1}{\sr{\cN}} \sum_{1\leq j_1 < j_2 \leq L} \cU_{j_1j_2}|j_1j_2\rag,
\ee
with the Bethe numbers $I_1,I_2$ which are integers and satisfy $0\leq I_1\leq I_2\leq L-1$.
There is
\be
\cU_{j_1j_2} = \ep^{\ii(j_1p_1+j_2p_2+\f\th2)} + \ep^{\ii(j_1p_2+j_2p_1-\f\th2)},
\ee
with $p_1,p_2,\th$ being solutions to the equation
\be \label{Betheequation}
\ep^{\ii\th} = - \f{1+\ep^{\ii(p_1+p_2)}-2\ep^{\ii p_1}}{1+\ep^{\ii(p_1+p_2)}-2\ep^{\ii p_2}}.
\ee
The normalization factor is
\be
\cN=\sum_{1\leq j_1 < j_2 \leq L} |\cU_{j_1j_2}|^2.
\ee
In the state $|I_1I_2\rag$ there are two magnons with physical momenta $p_1,p_2$ and momenta $k_1,k_2$ related as
\be
p_1 = \f{2\pi k_1}{L}, ~ p_2 = \f{2\pi k_2}{L}.
\ee

To the equation (\ref{Betheequation}), there are three cases of solutions, namely case I solution, case II solutions, and case III solutions.
The case III solutions could be further classified into case IIIa solutions and case IIIb solutions.

\subsubsection{Case I solution}

For the case I solution, there are trivially
\be
I_1=p_1=k_1=I_2=p_2=k_2=\th=0,
\ee
and the state is
\be
|00\rag = \sr{\f{2}{L(L-1)}} \sum_{1\leq j_1<j_2\leq L} |j_1j_2\rag.
\ee
The probability of the local state $|j_1j_2\rag$ is
\be
p^\I_{j_1j_2}=\frac{2}{L(L-1)}, ~ 1 \leq j_1<j_2\leq L,
\ee
which is the same as the probability distribution (\ref{twoindpar}) in the configuration of two identical hard-core classical particles in subsection~\ref{sectionhard2ind}.

\subsubsection{Case II solution}

For the case II solution $|I_1I_2\rag$, there are
\be \label{k1k2fromI1I2theta}
k_1=I_1+\f{\th}{2\pi}, ~
k_2=I_2-\f{\th}{2\pi},
\ee
with Bethe numbers $I_1,I_2$ which satisfy $0\leq I_1 < I_2\leq L-1$ and real shift angle $\th\in[0,\pi]$.
We have
\be
k_{12}=I_{12}+\f{\th}{\pi},
\ee
with the momentum difference $k_{12}=k_1-k_2$ and Bethe number difference $I_{12}=I_1-I_2$.
The normalization factor is
\be
\cN = L(L-1) + \f{L\cos(\th-p_{12})-(L-1)\cos\th-\cos(\th-L p_{12})}{1-\cos p_{12}}.
\ee

The probability that one finds the total system in the local state $|j_1j_2\rag$ is
\be \label{XXXpIIj1j2}
p^\II_{j_1j_2} = \f{2}{\cN} [ 1+ \cos ( j_{12} p_{12} + \th ) ], ~ 1 \leq j_1 < j_2 \leq L,
\ee
with $j_{12}=j_1-j_2$ and $p_{12}=p_1-p_2$.
For the subsystem $A$, the probabilities of the subsystem local states $|G\rag_A$, $|j\rag_A$, and $|j_1j_2\rag_A$ are, respectively,
\bea
&& \hspace{-16mm} p^{A,\II}_0 = \f{1}{\cN} \Big[ (L-\ell)(L-\ell-1)
                       +\f{(L-\ell)\cos(p_{12}-\th)-(L-\ell-1)\cos\th-\cos[(L-\ell)p_{12}-\th]}{1-\cos p_{12}} \Big], \label{pAII0}\\
&& \hspace{-16mm} p^{A,\II}_j = \f{2}{\cN} \Big[ L-\ell + \f{\sin\f{p_{12}(L-\ell)}{2}\cos[p_{12}(j-\f{L+\ell+1}{2})+\th]}{\sin\f{p_{12}}{2}} \Big],
~ j \in [1,\ell], \label{pAIIj}\\
&& \hspace{-16mm} p^{A,\II}_{j_1j_2} = \f{2}{\cN} [ 1+ \cos ( j_{12} p_{12} + \th ) ], ~ 1 \leq j_1 < j_2 \leq \ell,\label{pAIIj1j2}
\eea
which could be obtained as the marginal probabilities of (\ref{XXXpIIj1j2}).

\subsubsection{Case IIIa solution}

For the case IIIa solution, there are Bethe numbers
\be
I_1=\f{I-1}{2}, ~
I_2=\f{I+1}{2},
\ee
with the total Bethe number
\be
I = \td I,\td I+2,\cdots,\f{L}{2}-1,\f{3L}{2}+1,\f{3L}{2}+3,\cdots,2L-\td I,
\ee
where $\td I$ is an odd integer around $2\sr{L}/\pi$ in $L\to+\inf$ limit.
Without loss of generality, we only need to consider $I = \td I,\td I+2,\cdots,\f{L}{2}-1$.
The solution to the Bethe equation is
\be \label{XXXIIIap1p2}
p_1=\f{\pi I}{L}+\ii v, ~
p_2=\f{\pi I}{L}-\ii v, ~
\th=\pi+\ii L v.
\ee
In $L\to+\inf$ limit, there is
\be
v = - \log \Big| \cos \f{\pi I}{L} \Big|,
\ee
which is in the range
\be
\f{2}{L} \lesssim v \lesssim \log\f{L}{\pi}.
\ee
One could understand $1/v$ as the size of the bound state.
The normalization factor is
\be
\cN = L \Big[ \f{\sinh[(L-1)v]}{\sinh v} - (L-1) \Big].
\ee

The probability that one finds the total system in the local state $|j_1j_2\rag$ is
\be\label{XXXpIIIaj1j2}
p^\IIIa_{j_1j_2} = \f{4}{\cN} \sinh^2 \Big[ v \Big( j_{12} + \f{L}{2} \Big) \Big], ~ 1 \leq j_1 < j_2 \leq L,
\ee
with $j_{12}=j_1-j_2$.
For the subsystem $A$, the probabilities of the subsystem local states $|G\rag_A$, $|j\rag_A$, and $|j_1j_2\rag_A$ are, respectively,
\bea
&& p^{A,\IIIa}_0 = \f{1}{\cN} \Big[ (L-\ell) \f{\sinh[v(L-1)]}{\sinh v}
                           - \f{\sinh(v\ell)\sinh[v(L-\ell)]}{\sinh^2v}
                           - (L-\ell)(L-\ell-1) \Big], \label{pAIIIa0} \\
&& p^{A,\IIIa}_j = \f{2}{\cN} \Big[ \f{\sinh[v(L-\ell)]\cosh[2v(j-\f{\ell+1}{2})]}{\sinh v}
                           - (L-\ell) \Big], ~
            j \in [1,\ell], \label{pAIIIaj} \\
&& p^{A,\IIIa}_{j_1j_2} = \f{4}{\cN} \sinh^2 \Big[ v \Big( j_{12} + \f{L}{2} \Big) \Big], ~ 1 \leq j_1 < j_2 \leq \ell, \label{pAIIIaj1j2}
\eea
which are just the marginal probabilities of (\ref{XXXpIIIaj1j2}).

For finite $v$ in the limit $L\to+\inf$, we have the probabilites
\be
p^\IIIa_{j_1j_2} \to \f{2\sinh v}{L} \ep^{ 2v ( |j_{12}+\f{L}{2}| - \f{L - 1}{2} ) }, ~ 1 \leq j_1 < j_2 \leq L,
\ee
and in the scaling limit there are
\bea
&& p^\IIIa_0 \to 1-x, \\
&& p^\IIIa_j \to \f{1}{L} \ep^{ - 2v ( \f{\ell - 1}{2} - |j-\f{\ell+1}{2}| ) }, ~ j \in [1,\ell], \\
&& p^\IIIa_{j_1j_2} \to \f{2\sinh v}{L} \ep^{ - 2v ( \f{L - 1}{2} - |j_{12}+\f{L}{2}| ) }, ~ 1 \leq j_1 < j_2 \leq \ell.
\eea

\subsubsection{Case IIIb solution}

For the case IIIb solution, there are Bethe numbers
\be
I_1=I_2=\f{I}{2},
\ee
with the total Bethe number
\be
I=2,4,\cdots,\f{L}{2},\f{3L}{2},\f{3L}{2}+2,\cdots,2L-2.
\ee
In the case with $I=\f{L}{2}$, the state is extremely bound
\be
|I_1I_2\rag = \f{1}{\sqrt{L}} \sum_{j=1}^L \ep^{\f{2\pi\ii I}{L}(j+\f12)}|j,j+1\rag,
\ee
the results are the same as the those in the single-magnon state in subsection~\ref{sectionXXXI}, and we will not repeat the calculations in this paper.
Without loss of generality, we only consider $I = 2,4,\cdots,\f{L}{2}-2$.
The solution to the equation (\ref{Betheequation}) is
\be \label{XXXIIIbp1p2}
p_1=\f{\pi I}{L}+\ii v, ~
p_2=\f{\pi I}{L}-\ii v, ~
\th=\ii L v.
\ee
In $L\to+\inf$ limit, there is still
\be
v = - \log \Big| \cos \f{\pi I}{L} \Big|,
\ee
which is in the range
\be
\f{2\pi^2}{L^2} \lesssim v \lesssim \log\f{L}{\pi}.
\ee
The normalization factor is
\be
\cN = L \Big[ \f{\sinh[(L-1)v]}{\sinh v} + (L-1) \Big].
\ee

The probability that one finds the total system in the local state $|j_1j_2\rag$ is
\be \label{XXXpIIIbj1j2}
p^\IIIb_{j_1j_2} = \f{4}{\cN} \cosh^2 \Big[ v \Big( j_{12} + \f{L}{2} \Big) \Big], ~ 1 \leq j_1 < j_2 \leq L,
\ee
with $j_{12}=j_1-j_2$.
For the subsystem $A$, the probabilities of the subsystem local states $|G\rag_A$, $|j\rag_A$, and $|j_1j_2\rag_A$ are, respectively,
\bea
&& p^{A,\IIIb}_0 = \f{1}{\cN} \Big[ (L-\ell) \f{\sinh[v(L-1)]}{\sinh v}
                           - \f{\sinh(v\ell)\sinh[v(L-\ell)]}{\sinh^2v}
                           + (L-\ell)(L-\ell-1) \Big], \label{pAIIIb0} \\
&& p^{A,\IIIb}_j = \f{2}{\cN} \Big[ \f{\sinh[v(L-\ell)]\cosh[2v(j-\f{\ell+1}{2})]}{\sinh v}
                           + (L-\ell) \Big], ~
            j \in [1,\ell], \label{pAIIIbj} \\
&& p^{A,\IIIb}_{j_1j_2} = \f{4}{\cN} \cosh^2 \Big[ v \Big( j_{12} + \f{L}{2} \Big) \Big], ~ 1 \leq j_1 < j_2 \leq \ell, \label{pAIIIbj1j2}
\eea
which are the marginal probabilities of (\ref{XXXpIIIbj1j2}).

\section{SSH model}\label{appendixSSH}

In this appendix, we consider the Su-Schrieffer-Heeger (SSH) model \cite{Su:1979ua,Su:1979wc} of a ring with $L$ unit cells
\be
H = \sum_{j=1}^L [ v ( a^\dag_{j,1}a_{j,2} + a^\dag_{j,2}a_{j,1} ) + u ( a^\dag_{j,2}a_{j+1,1} + a^\dag_{j+1,1}a_{j,2} ) ],
\ee
which describes electrons hopping on a dimerised lattice.
In each unit cell labeled by $j=1,2,\cdots,L$, there are fermionic modes $a_{j,\a},a^\dag_{j,\a}$ with $\a=1,2$.
We only consider periodic boundary conditions.
One could choose that the hopping constants $v$ inside each unit cell and $u$ between neighboring unit cells are nonnegative real numbers.

The SSH model could be diagonalized by the Fourier transformation
\bea
&& b^\dag_{k,\a} = \f{1}{\sqrt{L}} \sum_{j=1}^L \ep^{\f{2\pi\ii j k}{L}} a^\dag_{j,\a}, ~~
   b_{k,\a} = \f{1}{\sqrt{L}} \sum_{j=1}^L \ep^{-\f{2\pi\ii j k}{L}} a_{j,\a}, \nn\\
&& k=0,1,\cdots,L-1, ~ \a=1,2.
\eea
followed by the transformation
\be
c^\dag_{k,\pm} = \f{1}{\sqrt{2}} ( b^\dag_{k,1} \pm \ep^{\ii\th_k} b^\dag_{k,2} ),~~
c_{k,\pm} = \f{1}{\sqrt{2}} ( b_{k,1} \pm \ep^{-\ii\th_k} b_{k,2} ),
\ee
where the angle $\th_k$ is determined by
\be
\ep^{\ii\th_k}=\f{v+w\ep^{\f{2\pi\ii k}{L}}}{\ve_k}, ~~
\ve_k = \sqrt{v^2+u^2+2vu\cos\f{2\pi k}{L}}.
\ee
The Hamiltonian becomes that of the double copies of the free fermionic chains
\be
H = \sum_k \ve_k ( c^\dag_{k,+} c_{k,+} - c^\dag_{k,-} c_{k,-} ).
\ee

In the SSH model there are two sets of independent fermionic modes, namely the modes $c_{k,+},c^\dag_{k,+}$ and the modes $c_{k,-},c^\dag_{k,-}$.
One may define the empty state $|\ves\rag$ as
\be
c_{k,\pm} |\ves\rag = 0, ~ k=0,1,\cdots,L-1,
\ee
based on which a general excited states $|K_+,K_-\rag$ with modes $K_+=\{k^+_1,k^+_2,\cdots,k^+_{r_+}\}$ and $K_-=\{k^-_1,k^-_1,\cdots,k^-_{r_-}\}$ may be constructed as
\be
|K_+,K_-\rag =
c^\dag_{k^+_1,+}c^\dag_{k^+_2,+}\cdots c^\dag_{k^+_{r_+},+}
c^\dag_{k^-_1,-}c^\dag_{k^-_2,-}\cdots c^\dag_{k^-_{r_-},-}
|\ves\rag.
\ee
As the modes $K_+$ and $K_-$ are independent, their contributions to the entanglement entropy and Shannon entropy are also independent.
Then we have the entanglement entropy in the SSH model as the sum of the entanglement entropies in the free fermionic chain
\bea
S^\SSH_{A,\{K_+,K_-\}} = S^\fer_{A,K_+} + S^\fer_{A,K_-}.
\eea
There are similar results for the Shannon entropy and mutual information
\bea
&& H^\SSH_{\{K_+,K_-\}}(L) = H^\fer_{A,K_+}(L) + H^\fer_{A,K_-}(L), \nn\\
&& H^\SSH_{\{K_+,K_-\}}(\ell) = H^\fer_{A,K_+}(L) + H^\fer_{A,K_-}(\ell), \nn\\
&& M^\SSH_{\{K_+,K_-\}}(\ell) = M^\fer_{A,K_+}(L) + M^\fer_{A,K_-}(\ell).
\eea

\section{Classical particles} \label{appendixCP}

This appendix presents the calculation of the Shannon entropy and mutual information for classical particle configurations on a circular chain. Both soft-core and hard-core particles are considered, corresponding to the classical limits of bosonic and fermionic quantum particles, respectively.

\subsection{Soft-core classical particles}

We start with soft-core classical particles, which can have any number of particles at one site, i.e.\ a site can be empty, occupied by one particle, or occupied by more than one particle.

\subsubsection{One particle}\label{sectionsoft1}

We consider the macroscopic configuration of the chain in which there is one soft-core classical particle.
There are $L$ possible microscopic configurations, i.e.\ the  particle at site $j$ with $j \in [1,L]$, and probabilities are
\be
p_j = \f{1}{L}, ~ j \in [1,L].
\ee
The Shannon entropy of the whole system is
\be
H^\soft_1(L)= \log{L}.
\ee

For the subsystem $A$, there are $\ell+1$ possible microscopic configurations, as shown in Table~\ref{tablesoft1sub}.
The Shannon entropy of the subsystem $A$ is
\be
H^\soft_1(\ell) = x \log L - (1-x) \log (1-x).
\ee
The Shannon mutual information is
\be
M^\soft_1(\ell) = - x \log x - (1-x) \log (1-x),
\ee
which is just the Shannon entropy of the probability distribution of the coarse-grained subsystem configurations $\{x,1-x\}$.
Observe that the values of $1-x$ and $x$ correspond to the presence of no particle and one particle, respectively, in the subsystem $A$.

\begin{table}[h]
  \centering
\begin{tabular}{|c|c|c|c|} \hline
   microscopic configurations      & probabilities    & ranges      & numbers \\ \hline
   no particle in $A$              & $p_{A,0} = 1 - x$    & -           & 1       \\ \hline
   one particle in $A$ at $j$ & $p_{A,j} = \f{1}{L}$ & $j \in [1,\ell]$ & $\ell$  \\ \hline
\end{tabular}
  \caption{The microscopic configurations and probabilities of the subsystem $A$ in the macroscopic configuration of one soft-core classical particle.}\label{tablesoft1sub}
\end{table}

\subsubsection{Two identical particles}\label{sectionsoft2ind}

We consider the macroscopic configuration in which there are two identical particles on the chain.
For the total system, there are $\frac{L(L+1)}{2}$ possible microscopic configurations as shown in Table~\ref{tablesoft2indtot}.
In the scaling limit, the Shannon entropy of the total system is
\be
H^\soft_{1^2}(L) = 2\log L-\log 2 
.
\ee

\begin{table}[h]
  \centering
\begin{tabular}{|c|c|c|c|} \hline
   microscopic configurations          & probabilities             & ranges                    & numbers         \\ \hline
   both at $j$                         & $p_{j^2} = \f{1}{L^2}$    & $j \in [1,L]$         & $L$             \\ \hline
   one at $j_1$ and the other at $j_2$ & $p_{j_1j_2} = \f{2}{L^2}$ & $1 \leq j_1 < j_2 \leq L$ & $\f{L(L-1)}{2}$ \\ \hline
\end{tabular}
  \caption{The microscopic configurations and probabilities of the total system in the macroscopic configuration of two identical soft-core classical particles.}\label{tablesoft2indtot}
\end{table}

For the subsystem $A$, there are $\f{(\ell+1)(\ell+2)}{2}$ different microscopic configurations, as shown in Table~\ref{tablesoft2indsub}.
In the scaling limit, we get the Shannon entropy of the subsystem $A$
\be \label{Hsoft1hat2ell}
H^\soft_{1^2}(\ell) = 2x \log L - x(2-x) \log2 -2(1-x)\log(1-x) 
.
\ee
In the scaling limit, the mutual information is
\bea \label{Isoft1hat2ell}
&& M^\soft_{1^2}(\ell)=-x^2\log x^2- 2x(1-x)\log [2x(1-x)]-(1-x)^2\log (1-x)^2 
\nn\\
&& \phantom{M^\soft_{1^2}(\ell)}
   = -2x\log x -2(1-x)\log(1-x) -2x(1-x)\log2 
   ,
\eea
which is just the Shannon entropy of the probability distribution of the coarse-grained subsystem configurations $\{ x^2, 2x(1-x), (1-x)^2 \}$.

\begin{table}[h]
  \centering
\begin{tabular}{|c|c|c|c|} \hline
   microscopic configurations                           & probabilities           & ranges                       & numbers \\ \hline
   no particle in $A$                                   & $p_{A,0} = (1 - x)^2$       & -                            & 1       \\ \hline
   only one particle in $A$ at $j$                      & $p_{A,j} = \f{2(1-x)}{L}$   & $j \in [1,\ell]$             & $\ell$  \\ \hline
   both in $A$ at $j$                                   & $p_{A,j^2} = \f{1}{L^2}$  & $j \in [1,\ell]$             & $\ell$  \\ \hline
   both in $A$ with one at $j_1$ and the other at $j_2$ & $p_{A,j_1j_2}=\f{2}{L^2}$ & $1 \leq j_1 < j_2 \leq \ell$ & $\f{\ell(\ell-1)}{2}$ \\ \hline
\end{tabular}
  \caption{The microscopic configurations and probabilities of the subsystem $A$ in the macroscopic configuration of two identical soft-core classical particles.}\label{tablesoft2indsub}
\end{table}

\subsubsection{Two distinguishable particle}\label{sectionsoft2dis}

We consider two distinguishable particles, say one red particle and one blue particle.
There are $L^2$ possible microscopic configurations, as shown in Table~\ref{tablesoft2distot}.
The Shannon entropy is
\be
H^\soft_{12}(L) = 2 \log L.
\ee

\begin{table}[h]
  \centering
\begin{tabular}{|c|c|c|c|} \hline
   microscopic configurations      & probabilities             & ranges                                & numbers  \\ \hline
   both at $j$                     & $p_{j^2} = \f{1}{L^2}$    & $j \in [1,L]$                     & $L$      \\ \hline
   red at $j_1$ with blue at $j_2$ & $p_{j_1j_2} = \f{1}{L^2}$ & $j_1,j_2\in[1,L]$, $j_1 \neq j_2$ & $L(L-1)$ \\ \hline
\end{tabular}
  \caption{The microscopic configurations and probabilities of the total system in the macroscopic configuration of two distinguishable soft-core classical particles.}\label{tablesoft2distot}
\end{table}

For the subsystem $A$, there are $(\ell+1)(\ell+2)$ possible configurations, as shown in Table~\ref{tablesoft2dissub}.
We get the Shannon entropy of the subsystem
\be \label{Hsoft12ell}
H^\soft_{12}(\ell) = 2x\log L - 2(1-x)\log(1-x).
\ee
The mutual information is
\be \label{Isoft12ell}
M^\soft_{12}(\ell) = 2 [ - x \log x - (1-x)\log(1-x) ],
\ee
which is just the Shannon entropy of the probability distribution of the coarse-grained subsystem configurations $\{x,1-x\} \otimes \{x,1-x\} = \{ x^2, x(1-x),  x(1-x), (1-x)^2 \}$.

\begin{table}[h]
  \centering
\begin{tabular}{|c|c|c|c|} \hline
   microscopic configurations                   & probabilities                 & ranges                                   & numbers \\ \hline
   no particle in $A$                           & $p_{A,0} = (1 - x)^2$             & -                                        & 1       \\ \hline
   only red particle in $A$ at $j$              & $p_{A,j}^{\rm red} = \f{1-x}{L}$  & $j \in [1,\ell]$                         & $\ell$  \\ \hline
   only blue particle in $A$ at $j$             & $p_{A,j}^{\rm blue} = \f{1-x}{L}$ & $j \in [1,\ell]$                         & $\ell$  \\ \hline
   both in $A$ at $j$                           & $p_{A,j^2} = \f{1}{L^2}$          & $j =1,2,\cdots,L$                        & $\ell$  \\ \hline
   both in $A$ with red at $j_1$ and blue at $j_2$ & $p_{A,j_1j_2}=\f{1}{L^2}$      & $j_1,j_2\in[1,\ell]$, $j_1 \neq j_2$ & ${\ell(\ell-1)}$ \\ \hline
\end{tabular}
  \caption{The microscopic configurations and probabilities of the subsystem $A$ in the macroscopic configuration of two distinguishable soft-core classical particles.}\label{tablesoft2dissub}
\end{table}

\subsection{Hard-core classical particles}

We next consider classical particles with the restriction that a site can have only one particle at most.

\subsubsection{One particle} \label{sectionhard1}

This case of one hard-core classical particle is identical to that of one soft-core classical particle in subsection~\ref{sectionsoft1}. We will skip it here.


\subsubsection{Two identical particles} \label{sectionhard2ind}

We consider two identical hard-core classical particles.
There are $\f{L(L-1)}{2}$ possible microscopic configurations, i.e.\ that one particle at site $j_1$ and the other particle at site $j_2$ with $1\leq j_1 < j_2 \leq L$, and the corresponding probabilities are
\be \label{twoindpar}
p_{j_1j_2} = \f{2}{L(L-1)}, ~1\leq j_1 < j_2 \leq L.
\ee
In the scaling limit, the Shannon entropy of the total system is
\be \label{Hhard1hat2L}
H^\hard_{1^2}(L) = 2\log L-\log 2 
.
\ee

For the subsystem $A$, there are $\f{\ell(\ell+1)}{2}+1$ possible configurations, as shown in Table~\ref{tablehard2indsub}.
In the scaling limit, we get the Shannon entropy of the subsystem
\be \label{Hhard1hat2ell}
H^\hard_{1^2}(\ell) = 2x \log L -2(1-x)\log(1-x) - x(2-x) \log2 
.
\ee
In the scaling limit, the mutual information is
\bea \label{Ihard1hat2ell}
&& M^\hard_{1^2}(\ell)= -2x\log x -2(1-x)\log(1-x) -2x(1-x)\log2 
 \nn\\
&& \phantom{M^\hard_{1^2}(\ell)}
                   =-x^2\log x^2- 2x(1-x)\log [2x(1-x)]-(1-x)^2\log (1-x)^2 
                   ,
\eea
which is just the Shannon entropy of the probability distribution $\{x^2,2x(1-x),(1-x)^2\}$.
In the scaling limit, the results in this subsection are the same as those in subsection~\ref{sectionsoft2ind}.

\begin{table}[h]
  \centering
\begin{center}
\begin{tabular}{|c|c|c|c|} \hline
   microscopic configurations       & probabilities                          & ranges                      & numbers \\ \hline
   no particle in $A$               & $p_{A,0} = \f{(L-\ell)(L-\ell-1)}{L(L-1)}$ & -                           & 1       \\ \hline
   only one particle in $A$ at $j$       & $p_{A,j} = \f{2(L-\ell)}{L(L-1)}$          & $j \in [1,\ell]$            & $\ell$  \\ \hline
   both in $A$ with at $j_1$, $j_2$ & $p_{A,j_1j_2}=\f{2}{L(L-1)}$             & $1\leq j_1 < j_2 \leq \ell$ & $\f{\ell(\ell-1)}{2}$ \\ \hline
\end{tabular}
\end{center}
  \caption{The microscopic configurations and probabilities of the subsystem $A$ in the macroscopic configuration of two identical hard-core classical particles.}\label{tablehard2indsub}
\end{table}

\subsubsection{Two distinguishable particles}

We consider two distinguishable particles, say one red particle and one blue particle. There are $L(L-1)$ possible microscopic configurations, i.e.\ that the red particle at site $j_1$ and the blue particle at site $j_2$ with $j_1,j_2\in[1,L]$ and $j_1 \neq j_2$, and the corresponding probabilities are
\be
p_{j_1j_2} = \f{1}{L(L-1)}, ~ j_1,j_2\in[1,L], ~ j_1 \neq j_2.
\ee
In the scaling limit, the Shannon entropy is
\be
H^\hard_{12}(L) = 2 \log L 
.
\ee

For the subsystem $A$, there are $\ell^2+\ell+1$ possible configurations, as shown in Table~\ref{tablehard2dissub}.
In the scaling limit, we get the Shannon entropy of the subsystem
\be
H^\hard_{12}(\ell) = 2x\log L - 2(1-x)\log(1-x) 
.
\ee
In the scaling limit, the mutual information is
\be
M^\hard_{12}(\ell) = 2 [ - x \log x - (1-x)\log(1-x) ]
,
\ee
which is just the Shannon entropy of the probability distribution $\{x,1-x\} \otimes \{x,1-x\} = \{ x^2, x(1-x),  x(1-x), (1-x)^2 \}$.
In the scaling limit, the results in this subsection are the same as those in subsection~\ref{sectionsoft2dis}.

\begin{table}[h]
  \centering
\begin{center}
\begin{tabular}{|c|c|c|c|} \hline
   microscopic configurations                   & probabilities           & ranges                                   & numbers \\ \hline
   no particle in $A$                           & $p_{A,0} = \f{(L-\ell)(L-\ell-1)}{L(L-1)}$ & -                                        & 1       \\ \hline
  only red particle in $A$ at $j$                   & $p_{A,j}^{\rm red} = \f{L-\ell}{L(L-1)}$   & $j \in [1,\ell]$                     & $\ell$  \\ \hline
  only blue particle in $A$ at $j$                  & $p_{A,j}^{\rm blue} = \f{L-\ell}{L(L-1)}$  & $j \in [1,\ell]$                     & $\ell$  \\ \hline
   both in $A$ with red at $j_1$ and blue at $j_2$ & $p_{A,j_1j_2}=\f{1}{L(L-1)}$ & $j_1,j_2\in[1,\ell]$, $j_1 \neq j_2$ & ${\ell(\ell-1)}$ \\ \hline
\end{tabular}
\end{center}
  \caption{The microscopic configurations and probabilities of the subsystem $A$ in the macroscopic configuration of two distinguishable hard-core classical particles.}\label{tablehard2dissub}
\end{table}

\subsection{Summary and generalization}

When the number of particles is finite, the Shannon entropy and mutual information in the scaling limit only depend on the number of particles and how distinguishable they are, not on the limit of the number of particles at each site.
This means that we get the same results for soft-core and hard-core classical particles in the scaling limit. Moreover, the Shannon mutual information is equal to the Shannon entropy of the probability distribution of the coarse-grained subsystem configurations.

We summarize the results in this appendix as follows.
For the macroscopic configuration of one classical particle, in the scaling limit we get the total system Shannon entropy, subsystem Shannon entropy and mutual information
\bea
&& H^\cl_1(L)= \log{L}, \label{Hcl1L} \\
&& H^\cl_1(\ell) = x \log L - (1-x) \log (1-x), \label{Hcl1ell} \\
&& M^\cl_1(\ell) = - x \log x - (1-x) \log (1-x). \label{Icl1ell}
\eea
For two identical classical particles, we get
\bea
&& H^\cl_{1^2}(L) = 2\log L-\log 2,\label{Hcl1hat2L} \\
&& H^\cl_{1^2}(\ell) = 2x \log L -2(1-x)\log(1-x) - x(2-x) \log2,\label{Hcl1hat2ell}\\
&& M^\cl_{1^2}(\ell) = -x^2\log x^2- 2x(1-x)\log [2x(1-x)]-(1-x)^2\log (1-x)^2. \label{Icl1hat2ell}
\eea
For two distinguishable classical particles, we get
\bea
&& H^\cl_{12}(L) = 2 \log L,\label{Hcl12L}\\
&& H^\cl_{12}(\ell) = 2x\log L - 2(1-x)\log(1-x),\label{Hcl12ell}\\
&& M^\cl_{12}(\ell) = - 2 x \log x - 2(1-x)\log(1-x).\label{Icl12ell}
\eea

For more general cases, we just show the final results without giving any calculation details.
For $r$ identical particles with finite $r$ in the scaling limit $L\to+\inf$, $\ell\to+\inf$, and fixed $x={\ell}/{L}$, we obtain
\bea
&& H^\cl_{1^r}(L) = r \log L - \log r!, \label{Hcl1hatrL} \\
&& H^\cl_{1^r}(\ell) = r x \log L - \sum_{i=0}^r C_r^i x^i (1-x)^{r-i} \log [ i! C_r^i (1-x)^{r-i} ], \label{Hcl1hatrell} \\
&& M^\cl_{1^r}(\ell) = - \sum_{i=0}^r C_r^i x^i (1-x)^{r-i} \log [ C_r^i x^i (1-x)^{r-i} ], \label{Icl1hatrell}
\eea
with the binomial coefficient $C_r^i\equiv\f{r!}{i!(r-i)!}$.
For $s$ different kinds of particles with the number of the $i$-th kind of particles being $r_i$, $i=1,2,\cdots,s$ and total number of particles $R = \sum_{i=1}^s r_i$ being finite in the scaling limit, we get
\bea
&& H^\cl_{1^{r_1}2^{r_2}\cdots s^{r_s}}(L) = \sum_{i=1}^s H^\cl_{1^{r_i}}(L), \\
&& H^\cl_{1^{r_1}2^{r_2}\cdots s^{r_s}}(\ell) = \sum_{i=1}^s H^\cl_{1^{r_i}}(\ell), \label{Hcl1r12r2cdotssr2ell}\\
&& M^\cl_{1^{r_1}2^{r_2}\cdots s^{r_s}}(\ell) = \sum_{i=1}^s M^\cl_{1^{r_i}}(\ell).
\eea

In the scaling limit, the contributions of different classical particles to the total system Shannon entropy, subsystem Shannon entropy and mutual information become independent. However, the Shannon entropy and mutual information of quantum quasiparticles do not exhibit such decoupling property even in the limit of large momentum difference, as shown in the main text of the paper.

\section{Shannon entropy of particle number probability distribution} \label{appendixPNPD}

In this section, we determine the Shannon entropy for the probability distribution of subsystem particle numbers in free bosonic and fermionic chains, which are just the number entropy studied in for example \cite{Goldstein:2017bua,Xavier:2018twb,Capizzi:2022jpx}.
The semiclassical quasiparticle picture applies to the Shannon entropy of particle number probability distribution in the large momentum difference limit.

\subsection{Classical particles}

For finite number of classical particles in the scaling limit, the subsystem Shannon entropy of particle number probability distribution only depends on the total number of particles $R$, regardless the distinguishability of these particles.
The result is also not dependent on the limit of the particle number at one site.
In the limit $R \ll \ell<L$, the result is just the Shannon entropy of a binomial probability distribution
\be
H_R^\cl(\ell) = - \sum_{r=0}^R C_R^r x^r (1-x)^{R-r} \log[ C_R^r x^r (1-x)^{R-r} ],
\ee
with the binomial coefficient $C_R^r=\f{R!}{r!(R-r)!}$.

\subsection{Free bosonic chain}

In the single-particle state $|k\rag$ and double-particle state $|k^2\rag$ of the free bosonic chain, the results are trivial and the same as those of classical particles. We skip the calculations here.

In the double-particle state $|k_1k_2\rag$, the probabilities of finding zero, one, and two particles in the subsystem $A=[1,\ell]$ are respectively
\bea
&& p^\bos_{A,0} = p_{A,0}
                = (1 - x)^2 + \f{\sin^2(\pi k_{12} x)}{L^2\sin^2\f{\pi k_{12}}{L}}, \\
&& p^\bos_{A,1} = \sum_{j=1}^\ell p_{A,j}
                = 2 \Big[ x (1 - x) - \f{\sin^2(\pi k_{12} x)}{L^2\sin^2\f{\pi k_{12}}{L}} \Big], \\
&& p^\bos_{A,2} = \sum_{j=1}^\ell p_{A,j^2} + \sum_{1\leq j_1 <j_2\leq\ell} p_{A,j_1j_2}
                = x^2 + \f{\sin^2(\pi k_{12} x)}{L^2\sin^2\f{\pi k_{12}}{L}},
\eea
with $p_{A,0}$, $p_{A,j}$, $p_{A,j^2}$ and $p_{A,j_1j_2}$ being defined in Table~\ref{tablebosk1k2sub}.
The corresponding Shannon entropy is just
\be
H^\bos_{k_1k_2}(\ell) = - p^\bos_{A,0} \log p^\bos_{A,0} - p^\bos_{A,1} \log p^\bos_{A,1} - p^\bos_{A,2} \log p^\bos_{A,2}.
\ee
In the large momentum difference limit $|k_{12}| \gg 1$, the result approaches that of two classical particles
\be
\lim_{|k_{12}|\to+\inf} H^\bos_{k_1k_2}(\ell) = H_2^\cl(\ell).
\ee

\subsection{Free fermionic chain}

In the double-particle state $|k_1k_2\rag$ of the free fermionic chain, the probability of finding zero, one, and two particles in the subsystem $A=[1,\ell]$ are respectively
\bea
&& p^\fer_{A,0} = p_{A,0}
                = (1 - x)^2 - \f{\sin^2(\pi k_{12} x)}{L^2\sin^2\f{\pi k_{12}}{L}}, \\
&& p^\fer_{A,1} = \sum_{j=1}^\ell p_{A,j}
                = 2 \Big[ x (1 - x) + \f{\sin^2(\pi k_{12} x)}{L^2\sin^2\f{\pi k_{12}}{L}} \Big], \\
&& p^\fer_{A,2} = \sum_{1\leq j_1 <j_2\leq\ell} p_{A,j_1j_2}
                = x^2 - \f{\sin^2(\pi k_{12} x)}{L^2\sin^2\f{\pi k_{12}}{L}}.
\eea
with $p_{A,0}$, $p_{A,j}$ and $p_{A,j_1j_2}$ being defined in Table~\ref{tableferk1k2sub}.
The corresponding Shannon entropy is
\be
H^\fer_{k_1k_2}(\ell) = - p^\fer_{A,0} \log p^\fer_{A,0} - p^\fer_{A,1} \log p^\fer_{A,1} - p^\fer_{A,2} \log p^\fer_{A,2},
\ee
and in the large momentum difference limit $|k_{12}| \gg 1$ it also approaches that of two classical particles
\be
\lim_{|k_{12}|\to+\inf} H^\fer_{k_1k_2}(\ell) = H_2^\cl(\ell).
\ee

\section{Shannon entropy in $\sigma_j^x$ basis of XXX chain} \label{appendixXB}

This appendix focuses on the investigation of the Shannon entropy in the local basis of eigenstates of $\sigma_j^x$ for the XXX chain.%
\footnote{We thank M.~A.~Rajabpour for suggesting us to look into the case of Shannon entropy in the $\s_j^x$ local basis.}

\subsection{The $\sigma_j^x$ basis}

For a single set of Pauli matrices $\{\s^x,\s^y,\s^z\}$, the eigenstates of $\s^x$ is
\be
|\pm\rag = \f{1}{\sqrt{2}}( |\!\uparrow\rag \pm |\!\downarrow\rag ).
\ee

In the local basis of $\s_j^x$ eigenstates, there are local states which we denote as $|\cX\rag$ with $\cX$ being a set of numbers in the range $[1,L]$. In the state $|\cX\rag$, the site $j\notin\cX$ has spin $|+\rag$ and the site $j\in\cX$ has spin $|-\rag$.
For example, in this appendix we use the local states
\bea
&& |\ves\rag = |++\cdots+\rag, \nn\\
&& |j\rag = |\cdots-_j\cdots\rag, \nn\\
&& |j_1j_2\rag = |\cdots-_{j_1}\cdots -_{j_2}\cdots\rag,
\eea
where all the omitted sites have spin $|+\rag$.
For the total system with $L$ sites, there are $2^L$ states $\{|\cX\rag\}$ in $\s_j^x$ basis.
For the subsystem $A=[1,\ell]$, there are similarly $2^\ell$ states $\{|\cX\rag_A\}$.

\subsection{Ground state}

In the ground state of the ferromagnetic XXX chain $|G\rag = |\!\uparrow\uparrow\cdots\uparrow\rag$, the probability of the local state $|\cX\rag$ is
\be
p_\cX^G = \f{1}{2^L},
\ee
from which we get the Shannon entropy of the total system
\be
H^\XXX_G(L) = L \log 2.
\ee
Similarly, we get the probability of the local state $|\cX\rag_A$
\be
p_\cX^{A,G} = \f{1}{2^\ell},
\ee
and the subsystem Shannon entropy
\be
H^\XXX_G(\ell) = \ell \log 2.
\ee
Note that the total system and subsystem Shannon entropies take the maximal values.
The subsystem Shannon mutual information is trivial
\be
M^\XXX_G(\ell) = 0.
\ee

\subsection{Single-magnon state}

In the single-magnon state
\be
|I\rag = \f{1}{\sqrt{L}} \sum_{j=1}^L \ep^{\f{2\pi\ii j I}{L}} |j\rag, ~ I=0,1,\cdots,L-1,
\ee
the probability of the local state $|\cX\rag$ of the total system is
\be
p_\cX^I = \f{1}{2^L L} \Big| \sum_{j=1}^L \ep^{\f{2\pi\ii j I}{L}} m_{j,\cX} \Big|^2,
\ee
with $m_{j,\cX}=1$ for $j\notin\cX$ and $m_{j,\cX}=-1$ for $j\in\cX$.
The total system Shannon entropy could be evaluated numerically as
\be \label{XBtotRE}
H^\XXX_I(L) = - \sum_\cX p_\cX^I \log p_\cX^I.
\ee
Similarly, the probability of the subsystem local state $|\cX\rag_A$ is
\be
p_\cX^{A,I} = \f{1}{2^\ell} \Big( 1 - \f{\ell}{L} + \f{1}{L} \Big| \sum_{j=1}^\ell \ep^{\f{2\pi\ii j I}{L}} m_{j,\cX} \Big|^2 \Big),
\ee
from which we calculate the subsystem Shannon entropy as
\be \label{XBsubRE}
H^\XXX_I(\ell) = - \sum_\cX p_\cX^{A,I} \log p_\cX^{A,I}.
\ee
The subsystem Shannon mutual information is
\be \label{XBsubMI}
M^\XXX_I(\ell) = H^\XXX_I(\ell) + H^\XXX_I(L-\ell) - H^\XXX_I(L).
\ee

\begin{figure}[t]
  \centering
  \includegraphics[height=0.28\textwidth]{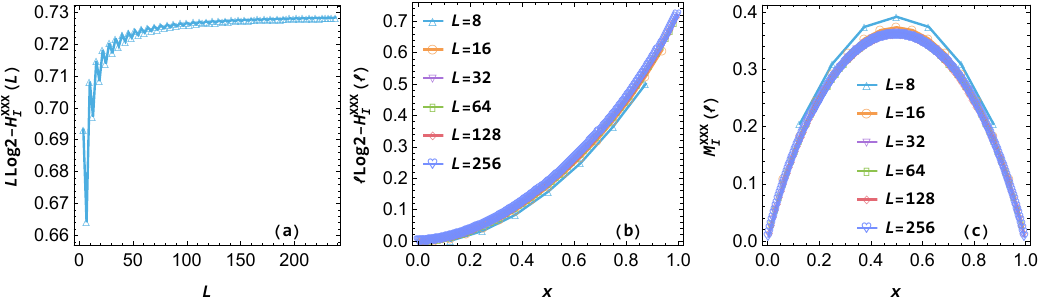}\\
  \caption{The $\sigma_j^x$ basis total system and subsystem Shannon entropies and mutual information in the single-magnon state of the XXX chain with Bethe numbers $I=0$ and $I=L/2$.}
  \label{FigureXBIeq0}
\end{figure}

\begin{figure}[t]
  \centering
  \includegraphics[height=0.28\textwidth]{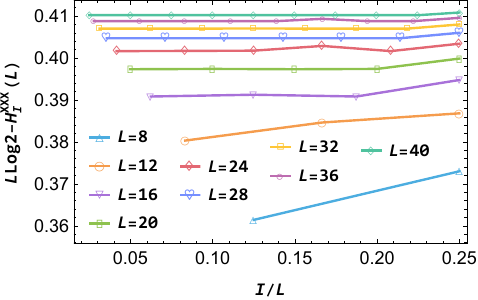}\\
  \caption{The $\sigma_j^x$ basis total system Shannon entropy $H^\XXX_I(L)$ (\ref{XBtotRE}) in the single-magnon state of the XXX chain for $I\in[1,L/4]$.}
  \label{FigureXBtotRE}
\end{figure}

For the special cases $I=0$ and $I=L/2$, the formulas of the Shannon entropies $H^\XXX_I(L)$ (\ref{XBtotRE}) and $H^\XXX_I(\ell)$ (\ref{XBsubRE}) could be further simplified, and we obtain
\bea
&& \hspace{-6mm} H^\XXX_0(L) = H^\XXX_{L/2}(L) = L \log 2 - \f{1}{2^L} \sum_{n=0}^L C_L^n \Big( L-4n+\f{4n^2}{L} \Big) \log \Big( L-4n+\f{4n^2}{L} \Big), \\
&& \hspace{-6mm} M^\XXX_0(L) = M^\XXX_{L/2}(L) = \ell \log 2 \nn\\
&& \hspace{-6mm} ~~~~~~~~~~~~~~~~~~~
- \f{1}{2^\ell} \sum_{n=0}^\ell C_\ell^n \Big( 1 + \f{\ell^2-(4n+1)\ell+4n^2}{L} \Big) \log \Big( 1 + \f{\ell^2-(4n+1)\ell+4n^2}{L} \Big).
\eea
We show the special results of the total system and subsystem Shannon entropies and mutual information with $I=0$ and $I=L/2$ in Figure~\ref{FigureXBIeq0}.
From the left panel we see that in large $L$ limit there are
\be
H^\XXX_0(L) = H^\XXX_{L/2}(L) = L \log 2 - C_{0} = L \log 2 - C_{L/2},
\ee
with the $L$-independent constant $C_{0}=C_{L/2}\approx0.73$
From the middle and right panels, we see that in the scaling limit there are
\bea
&& H^\XXX_0(\ell) = \ell \log 2 - F_0(x) = H^\XXX_{L/2}(\ell) = \ell \log 2 - F_{L/2}(x), \\
&& M^\XXX_0(\ell) = G_0(x)= M^\XXX_{L/2}(\ell) = G_{L/2}(x),
\eea
with the finite functions $F_0(x)=F_{L/2}(x)$ and $G_0(x)=G_{L/2}(x)$ depending on the ratio $x={\ell}/{L}$.

\begin{figure}[t]
  \centering
  \includegraphics[height=0.28\textwidth]{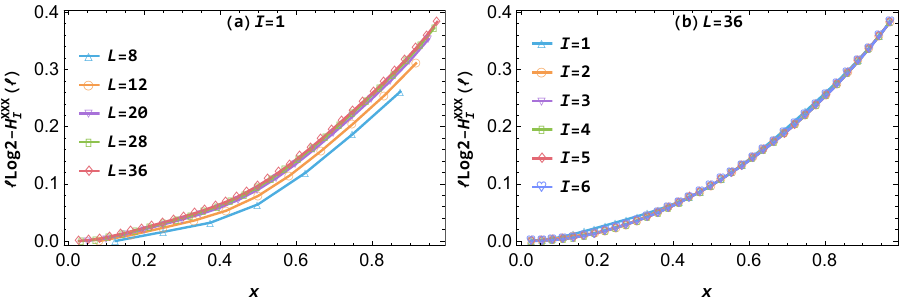}\\
  \caption{The $\sigma_j^x$ basis subsystem Shannon entropy $H^\XXX_I(\ell)$ (\ref{XBsubRE}) in the single-magnon state of the XXX chain.}
  \label{FigureXBsubRE}
\end{figure}

\begin{figure}[t]
  \centering
  \includegraphics[height=0.28\textwidth]{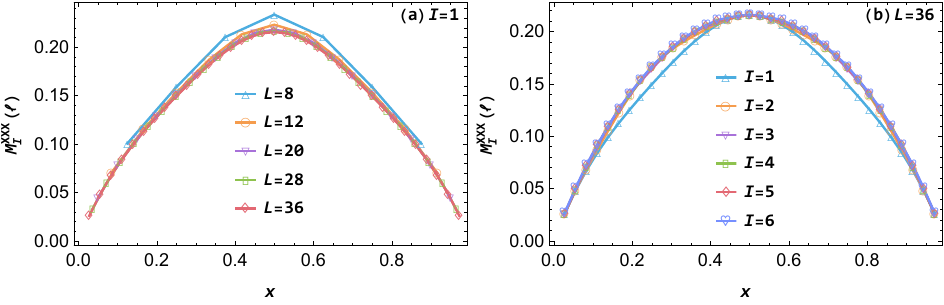}\\
  \caption{The $\sigma_j^x$ basis subsystem Shannon mutual information $M^\XXX_I(\ell)$ (\ref{XBsubMI}) in the single-magnon state of the XXX chain.}
  \label{FigureXBMI}
\end{figure}

For general $I$, the evaluations of the $\sigma_j^x$ basis total system Shannon entropy and the subsystem Shannon entropy and mutual information are exponentially difficult problems, and we could only numerically calculate them for not so large $L$, say $L\leq36$.
It is easy to see that both the total system and subsystem Shannon entropies $H^\XXX_I(L)$ (\ref{XBtotRE}) and $H^\XXX_I(\ell)$ (\ref{XBsubRE}) are invariant under the changes $I \to L-I$ and $I \to \f{L}{2}-I$, and so without loss of generality we only need to consider $I$ in the range $[1,{L}/{4}]$.
We show the numerical results of $H^\XXX_I(L)$ (\ref{XBtotRE}), $H^\XXX_I(\ell)$ (\ref{XBsubRE}), and $M^\XXX_I(\ell)$ (\ref{XBsubMI}) in, respectively figures~\ref{FigureXBtotRE}, \ref{FigureXBsubRE} and \ref{FigureXBMI}.
From Figure~\ref{FigureXBtotRE}, we see that in large $L$ limit the total system Shannon entropy takes the form
\be
H^\XXX_I(L) = L \log2 - C_I,
\ee
with $C_I$ being a constant that depends on the relative values of $I$ and $L$. We anticipate that for most values of $I$ there is the universal constant $C_I\approx0.41$, and for a few exception values of $I$ the constant $C_I$ may take some exceptional values, like the cases $I=0$ and $I=L/2$ discussed above.
From Figure~\ref{FigureXBsubRE}, in the scaling limit the subsystem Shannon entropy takes the form
\be
H^\XXX_I(\ell) = \ell \log2 - F_I(x),
\ee
with the $I$-dependence function $F_I(x)$.
From Figure~\ref{FigureXBMI}, in the scaling limit the subsystem Shannon mutual information takes the form
\be
M^\XXX_I(\ell) = G_I(x),
\ee
with the $I$-dependence function $G_I(x)$.
We expect that in the scaling limit, for $I\gg1$ with a few possible exceptional values excluded, the functions $F_I(x)$ and $G_I(x)$ will take on respective universal forms.

\providecommand{\href}[2]{#2}\begingroup\raggedright\endgroup


\end{document}